\documentclass{elsarticle}

\usepackage{lmodern}
\usepackage[T1]{fontenc}
\usepackage{amsmath}
\usepackage{amsthm}
\usepackage{nccmath}
\usepackage{boxedminipage}
\usepackage{amsfonts}
\usepackage{fancyhdr}
\usepackage{graphicx}
\usepackage{geometry}
\usepackage{epstopdf}
\usepackage{url}
\usepackage{hyperref}
\usepackage[nameinlink]{cleveref}
\usepackage{color}
\usepackage{booktabs}
\usepackage{multirow} 
\usepackage[shortlabels]{enumitem}
\usepackage{algorithm,algpseudocode}

\usepackage{tikz}
\usetikzlibrary{3d}
\usetikzlibrary{patterns}
\usetikzlibrary{calc}
\usetikzlibrary{arrows}

\usepackage{pgfplots}
\usepackage{pgfplotstable}
\pgfplotsset{compat=1.18}
\usepgfplotslibrary{groupplots}

\newtheorem{definition}{Definition}
\newtheorem{lemma}{Lemma}

\newtheorem{prop}{Property}
\newtheorem{cor}{Corollary}

\crefname{section}{Sec.}{Secs.}
\Crefname{section}{Section}{Sections}
\crefname{lemma}{Lem.}{Lems.}
\Crefname{lemma}{Lemma}{Lemmas}
\crefname{prop}{Prop.}{Props.}
\Crefname{prop}{Property}{Properties}
\crefname{cor}{Cor.}{Cors.}
\Crefname{cor}{Corollary}{Corollaries}
\crefname{definition}{Def.}{Defs.}
\Crefname{definition}{Definition}{Definitions}
\crefname{algorithm}{Alg.}{Algs.}
\Crefname{algorithm}{Algorithm}{Algorithms}
\crefname{figure}{Fig.}{Figs.}
\Crefname{figure}{Figure}{Figures}
\crefname{table}{Tab.}{Tabs.}
\Crefname{table}{Table}{Tables}

\definecolor{BoxCol}{rgb}{0.6196078,0.494117647,0.21960784}
\definecolor{wfugold}{rgb}{0.6196078,0.494117647,0.21960784}
\definecolor{wfubl}{rgb}{0.265625,0.41015625,0.48828125}
\definecolor{wfugr}{rgb}{0.33203125,0.4609375,0.1875}
\definecolor{wfur}{rgb}{0.59375,0.1953125,0.1328125}
\definecolor{wfublgr}{rgb}{0.296875,0.43359375,0.3359375}
\definecolor{wfup}{rgb}{0.3984375,0.125,0.2734375}

\colorlet{wfupL}{wfup!80!white}
\colorlet{wfupVL}{wfup!50!white}
\colorlet{wfurL}{wfur!75!white}

\usepackage{tensors}

\newcommand{\vc}[1]{\text{\normalfont vec}\left(#1\right)}
\newcommand{\Linear}{\mathbb{L}}
\newcommand{\Reverse}{\Linear^*}

\newcommand{\R}{\mathbb{R}}
\newcommand{\K}[3]{\left[\!\left[\M{#1}, \M{#2}, \M{#3}\right]\!\right]}

\let\M\Mx
\let\V\Vc

\newcommand{\va}{\V{a}}
\newcommand{\vb}{\V{b}}
\newcommand{\vcc}{\V{c}}
\newcommand{\vha}{\V{\hat{a}}}
\newcommand{\vhb}{\V{\hat{b}}}
\newcommand{\vhc}{\V{\hat{c}}}

\newcommand{\vr}{\V{r}}
\newcommand{\vg}{\V{g}}
\newcommand{\vs}{\V{s}}
\newcommand{\vv}{\V{v}}

\newcommand{\MJ}{\M{J}}
\newcommand{\MH}{\M{H}}
\newcommand{\MK}{\M{K}}
\newcommand{\FD}[2]{\frac{\partial #1}{\partial #2}}
\newcommand{\SD}[3]{\frac{\partial^2 #1}{\partial #2 \, \partial #3}}

\def\A{\M{A}}
\def\B{\M{B}}
\def\C{\M{C}}

\makeatletter
\renewcommand*\env@matrix[1][\arraystretch]{%
  \edef\arraystretch{#1}%
  \hskip -\arraycolsep
  \let\@ifnextchar\new@ifnextchar
  \array{*\c@MaxMatrixCols c}}
\makeatother

\pgfplotsset{
 cpaxis/.style={
   width=0.31\linewidth,
   height=4.5cm,
   grid=both,
   minor grid style={gray!15},
   major grid style={gray!30},
   tick label style={font=\tiny},
   yticklabel style={font=\tiny},
   title style={font=\small, yshift=-2pt},
   every axis plot/.append style={line width=0.8pt},
 }
}

\pgfplotsset{filter discard warning=false}

\begin{document}

\begin{frontmatter}
\title{Newton's Method for CP Decomposition}

\author[1]{DuBose Tuller}
\ead{tullwd25@wfu.edu}
\author[1]{Grey Ballard}
\ead{ballard@wfu.edu}
\author[2]{Hussam Al Daas}
\ead{hussam.al-daas@stfc.ac.uk}
\author[2]{Nicholas I. M. Gould}
\ead{nick.gould@stfc.ac.uk}

\address[1]{Department of Computer Science, Wake Forest University, Winston-Salem, NC 27106 USA}
\address[2]{Scientific Computing Department, STFC, Rutherford Appleton Laboratory, Didcot OX11 0QX, UK}

\begin{abstract}
We present an algorithm for computing the CP tensor decomposition using Newton's method.
The problem of fitting a CP model to a data tensor with least squares loss can be solved by various gradient-based optimization techniques, including Gauss-Newton methods that exploit the tensor structure of the problem and iteratively solve the linear system involving the approximate Hessian matrix to determine the step direction at each iteration.
Gauss-Newton methods exhibit superlinear convergence when the residual is small but can otherwise struggle to converge when the approximation to the Hessian is not accurate.
We show that the Gauss-Newton approach can be extended to Newton's method, using the true Hessian to obtain fast convergence even in the case of large residuals.
In particular, we derive efficient means of applying the true Hessian to vectors for use within an iterative solver, we employ globalization and regularization techniques to make the Newton model robust to ill-conditioned and/or indefinite Hessians, and we propose hybrid use of Gauss-Newton and Newton steps to balance convergence behavior both far and near local minima.
Our numerical experiments use synthetic examples to show in which scenarios Gauss-Newton and other methods struggle to converge, and we demonstrate the superiority of the hybrid approach on two application data sets.
\end{abstract}
\end{frontmatter}

\section{Introduction}

The CP decomposition, known variously as canonical polyadic, CANDECOMP, PARAFAC, rank decomposition, or separated representation, is one of the most widely used low-rank representations of multidimensional data.
Given an input data tensor, a CP decomposition is an approximation of the data by a sum of rank-one tensors, where a rank-one tensor is an outer product of vectors.
We refer to the number of summands or components in the CP representation as the rank of the model. 
The (vector) factors of the rank-one tensors are collected into a set of factor matrices, one for each mode of the tensor.
CP is particularly useful for interpretation, as it is typically unique (up to ambiguities discussed below) and each rank-one component represents an underlying pattern that describes a multi-way relationship among the different modes of the multidimensional data.

However, the problem of computing a CP approximation using a least-squares loss function has many well-studied challenges.
Determining the right approximation rank is a difficult problem, so in practice many different values of rank must be attempted.
Even for a fixed rank $r$, the corresponding optimization problem is nonlinear and nonconvex.
There may not exist a best rank-$r$ CP approximation, and even if the best approximation exists, the problem may be ill-conditioned.
Furthermore, the scaling and permutation ambiguities in the CP representation can disrupt the convergence of optimization methods.

Many algorithms for solving the CP approximation problem have been developed and optimized, and no single method is clearly superior to all others (see \cite{tomasiComparisonAlgorithmsFitting2006}, for example).
The simplest method is called Alternating Least Squares (ALS) \cite{CC70,Harshman70}, which is a block coordinate descent method that alternates over updating each factor matrix while holding all other factor matrices fixed.
ALS is relatively cheap per outer iteration and can converge quickly for some problems but suffers from slow convergence in many other cases.
Gradient-based optimization methods update all factor matrices at once, and the evaluation of the gradient has roughly the same cost as an ALS iteration.
Quasi-Newton methods such as LBFGS can yield faster convergence than ALS at minimal added expense beyond evaluating the gradient (see \cite{ADK11}, for example).

Variants of the Gauss-Newton method, which involve an approximation of the Hessian using first-order derivatives, can work very well for CP with least-squares loss.
The CP model with the residual function yields a Jacobian that is highly structured, so the solve for the step direction at each Gauss-Newton outer iteration can be performed by an iterative method (typically, a Krylov method) with cheap inner-iteration cost relative to the gradient evaluation (see \cite{VL19}, for example).
In particular, neither the Jacobian nor the Gauss-Newton Hessian approximation is represented explicitly, and instead matrix-vector products are evaluated implicitly.
The Gauss-Newton approximation of the Hessian is particularly accurate when the residual is small, and for problems where the CP model is a very good approximation to the data, Gauss-Newton methods can exhibit superlinear convergence.

However, in cases of large residuals, the Gauss-Newton approximation of the Hessian is not  accurate, and the convergence of Gauss-Newton methods can slow down or stall.
Input tensor data coming from most applications is typically noisy, and the CP approximation problem seeks to separate the low-rank signal from the noise, so the magnitude of the residual depends on the signal-to-noise ratio.
The central contribution of this work is the demonstration that the true Hessian matrix can be implicitly and efficiently applied in an implementation of Newton's method that overcomes this barrier and yields superlinear convergence even in cases of large residuals (e.g., noisy data).
We show that the true Hessian has tensor structure that can be exploited similarly to its Gauss-Newton approximation.
With efficient evaluation of Hessian matrix-vector products using precomputed temporary values, the cost of an inner iteration of computing a Newton step direction is comparable to that of applying the Gauss-Newton approximation to a vector and cheap relative to the gradient evaluation.
The computational cost of precomputation, which is performed once per Newton iteration, is only 50\% more than the cost of computing the gradient.
This implies that the overall per-iteration cost of Newton's method for CP is comparable to other gradient-based algorithms, including Gauss-Newton and ALS.

Newton's method presents its own set of challenges, and previous considerations of using Newton's method for CP often cite its cost and/or lack of robustness \cite{Tomasi06,Paatero97} (see also \cite{KT10,EH12} for the related problem of CP rounding).
While fast convergence is achievable in the neighborhood of a local minimum, the true Hessian may be indefinite and generally can lead to erratic behavior when far from a solution.
We use two techniques to mitigate the loss of robustness of Newton's algorithm compared to Gauss-Newton.
First, we employ unique globalization and regularization techniques that overcome negative curvature present in the Newton, but not the Gauss-Newton model, and coping with this more severely regularized subproblem is more than simply solving a (perturbed) linear system.
In this case, an inner iteration using a nonlinear solver like GLRT \cite{CGT09}, \cite[Chaps.~9 \& 10]{CartGoulToin22} is required.
Second, we take a hybrid approach, relying on Gauss-Newton in early iterations and switching to Newton when the gradient is sufficiently small.
This hybrid approach achieves all the advantages of Gauss-Newton methods with added robustness to achieve fast convergence even in the case of large residuals.

Our numerical experiments in \cref{sec:experiments} illustrate the behavior predicted by the theory.
We use synthetic examples to vary the difficulty of problems in terms of ill-conditioning (using collinearity of factors), noise, and mis-estimations of rank.
The experiments show that ill-conditioning can lead to very slow convergence of ALS and that both Gauss-Newton and Newton exhibit superlinear convergence for problems with small residuals.
We show that Gauss-Newton's convergence deteriorates in the presence of large residuals due to high noise or underestimation of the rank, while Newton's superlinear convergence is robust to the size of the residual.
However, our experiments also show that Newton's method is often slower to locate the neighborhood of a solution, resulting in more iterations overall for easier problems.
Our hybrid approach achieves the fewest iterations, following the trajectory of Gauss-Newton when it succeeds and gaining the benefits of Newton when Gauss-Newton struggles.
We consider two application data sets that involve noisy data that confirm the observations from the synthetic data experiments.

\section{Background}
\label{sec:background}

\subsection{Tensor Notation}
\label{sec:background:tensor}

We use calligraphic notation for tensors (e.g., $\Tn{X}$), capital boldface for matrices (e.g., $\M{A}$), lowercase boldface for vectors (e.g., $\vv$), and regular font for scalars (e.g., $f$).
We focus on 3-way tensors in this work, but the methods generalize to the $d$-way case.
In general, we assume the input data tensor $\Tn{X}\in\R^{m\times n\times p}$, and we index into those dimensions using $i$, $j$, and $k$, respectively.

In computing a CP approximation of $\Tn{X}$, we consider the CP model as a Kruskal tensor $\Tn{T}$ of rank $r$ represented by three factor matrices $\A\in\R^{m\times r}$, $\B\in\R^{n\times r}$, and $\C\in\R^{p\times r}$, and we use $\ell$ to index over the rank $r$.
In shorthand, we write $\Tn{T} = \K{A}{B}{C}$, where 
$$t_{ijk} = \sum_{\ell=1}^r a_{i\ell}b_{j\ell}c_{k\ell}, \text{\quad or equivalently \quad} \Tn{T} = \sum_{\ell=1}^r \Vc{a}{\ell} \top \Vc{b}{\ell} \top \Vc{c}{\ell},$$
where $\top$ denotes the vector outer product and $\Vc{a}{\ell}$ is the $\ell$th column of $\A$, $\Vc{b}{\ell}$ is the $\ell$th column of $\B$, and $\Vc{c}{\ell}$ is the $\ell$th column of $\C$.
Kruskal tensors have inherent ambiguities: the columns of the factor matrices can be permuted and the factors of each vector outer product can be re-scaled \cite[Sec.~10.4]{TensorTextbook}.
That is, considering consistent columns
$\va$, $\vb$ and $\vcc$ of the factor matrices $\A$, $\B$ and $\C$, the term
$\va \top \vb \top \vcc$ in the Kruskal tensor is identical to
$\vha \top \vhb \top \vhc$, whenever $\vha = \alpha \va$, 
$\vhb = \beta \vb$ and $\vhc = \gamma \vcc$ for any three scalars satisfying $\alpha \beta \gamma = 1$. 

\subsection{Derivative Notation}

Let $f(\vv): \R^n \rightarrow \R$. 
Then the gradient $\Vc{g}(\vv)=\nabla f (\vv)$ of $f$ is the vector of partial derivatives with respect to each of the elements of $\vv$, with $[\Vc{g}(\vv)]_i = \FD{f}{v_i}$.
The Hessian $\M{H}(\vv)=\nabla^2 f(\vv)$ is the matrix of second derivatives with respect to pairs of the elements of $\vv$, with $[\M{H}(\vv)]_{ij} = \SD{f}{v_i}{v_j}$.

Let $\vr(\vv): \R^n \rightarrow \R^m$. 
Then the derivative $\M{J}(\vv)$ of $\vr$ is the Jacobian matrix containing the partial derivatives with respect to each variable of $\vv$ and element of the output of $\vr$, with $[\M{J}(\vv)]_{ij} = \FD{r_i}{v_j}$.
Note that we follow the convention of considering gradients as column vectors, which implies that the gradient of an element of $\vr(\vv)$ is the transpose of the corresponding row of the Jacobian.

\begin{definition}[Derivatives of Nonlinear Sum of Squares]
\label{def:leastsquarederivatives}
Let $\vr(\vv): \R^n \rightarrow \R^m$ and  $f(\vv): \R^n \rightarrow \R$ be such that 
$f(\vv) = \frac{1}{2}\|\vr(\vv)\|^2$. Then the gradient $\vg(\vv)$ of $f(\vv)$ is given by
$$\vg(\vv) \equiv {\nabla f\left ( \vv \right ) = \M{J}'(\vv)\Vc{r}(\vv),}$$
and the Hessian $\M{H}(\vv)$ of $f$ is given by
$$\M{H}(\vv) \equiv {\nabla ^{2} f\left ( \vv \right ) = \M{J}'(\vv)\M{J}(\vv) + \sum_{i=1}^{m}r_{i}\left ( \vv \right ) \cdot \nabla ^{2}r_{i}\left ( \V{\vv} \right ).}$$
We denote the second term of this matrix $\M{K}(\vv)$.
\end{definition}

\subsection{Special Matrix and Vector Operations}

In developing optimization methods for functions and variables that have matrix and tensor shapes, we rely on vectorizing the principal quantities to obtain standard forms.
There are two common ways to linearize matrices and tensors into a vector form.

\begin{definition}[Linearization Functions]
Let $i\in [m]$ and $j\in [n]$, then $\Linear:[m]\times[n] \rightarrow [mn]$ is the natural linearization function such that
$$\Linear(i,j;m,n) = i+(j-1)m,$$
and $\Reverse:[m]\times[n] \rightarrow [mn]$ is the reverse linearization function such that
$$\Reverse(i,j;m,n) = j+(i-1)n.$$
\end{definition}

The vec operator takes a matrix $\M{A}$ or a tensor $\Tn{X}$ and linearizes it into a vector according to natural ordering.

\begin{definition}[vec Operator]
\label{def:vec}
If $\M{A}\in \R^{m\times n}$, then $\vc{\M{A}} \in \R^{mn}$ where
$$[\vc{\M{A}}]_\alpha = a_{ij} \qtext{with} \alpha = \Linear(i,j;m,n) = i + \left ( j - 1 \right )m.$$
If $\Tn{X}\in \R^{m\times n\times p}$, then $\vc{\Tn{X}} \in \R^{mnp}$ where
$$[\vc{\Tn{X}}]_\alpha = x_{ijk} \qtext{with} \alpha = i + \left ( j - 1 \right )m + \left(k-1\right) mn.$$
\end{definition}

We can reshape a tensor into a matrix using modal unfoldings.
We use the notation $\Tm{X}{1}$ to denote the mode-1 unfolding, which stacks the mode-1 fibers as columns into a matrix with columns in natural order.
That is, $[\Tm{X}{1}]_{i\alpha} = x_{ijk}$ for $\alpha=\Linear(j,k;n,p)$.
Likewise, $[\Tm{X}{2}]_{j\beta} = x_{ijk}$ for $\beta=\Linear(i,k;m,p)$ and $[\Tm{X}{3}]_{k\gamma}=x_{ijk}$ for $\gamma=\Linear(i,j;m,n)$.
In particular, this implies $\vc{\Tn{X}}=\vc{\Tm{X}{1}}$.

The Kronecker, Khatri-Rao, and Hadamard products are central to the analysis of methods for computing CP decompositions.

\begin{definition}[Kronecker Product]
\label{def:Kron}
If $\M{A}\in \R^{m\times p}$ and $\M{B} \in \R^{n\times q}$, then $\M{A}\krn \M{B} \in \R^{mn\times pq}$ where
$$[\M{A}\krn \M{B}]_{\alpha\beta} = a_{ik}b_{j\ell} \qtext{with} \begin{cases} \alpha = \Reverse(i,j;m,n) = j + (i-1)n\\ \beta = \Reverse(k,\ell;p,q) = \ell + (k-1)q\end{cases}.$$
\end{definition}

If $a$ and $b$ are column vectors of length $m$ and $n$, then $a\krn b$ is a column vector of length $mn$.

\begin{definition}[Khatri-Rao Product]
\label{def:kr}
If $\M{A} \in \R^{m \times p}$ and $\M{B} \in \R^{n \times p}$, then $\M{A} \odot \M{B} \in \R^{mn \times p}$ such that
$$[\M{A} \odot \M{B}]_{\alpha \ell} = a_{i\ell} b_{j\ell} \qtext{with} \alpha = \Reverse(i,j; m,n) = j + (i-1)n.$$
\end{definition}

The Khatri-Rao Product is equivalently the Kronecker products of each of the corresponding columns in $\M{A}$ and $\M{B}$.

\begin{definition}[Hadamard Product]
If $\M{A} \in \R^{m \times n}$ and $\M{B} \in \R^{m \times n}$, then $\M{A} \had \M{B} \in \R^{m \times n}$ with
$$[\M{A} \had \M{B}]_{ij} = a_{ij} b_{ij}.$$
\end{definition}

\begin{definition}[Tensor-Times-Matrix Product {\cite[Def.~3.14]{TensorTextbook}}]
\label{def:ttm}
    The mode-$k$ tensor-times-matrix (TTM) product between a tensor $\Tn{X}$ and a matrix $\M{U}$ is written $\Tn{Y} = \Tn{X} \ttm \M{U}$ and is defined by 
    $$\M{Y}{(k)} = \M{U}\M{X}{(k)}.$$
\end{definition}

The following properties are crucial to deriving representations of these types of products of matrices that can be evaluated efficiently.

\begin{prop}[{\cite[Eq. (4)]{HS81}}]
\label{prop:krontimesvec}
For $\M{A}\in\R^{m\times p}$, $\M{B}\in\R^{n\times q}$, and $\M{X}\in\R^{p\times q}$, we have
$$( \M{B} \krn \M{A} ) \vc{\M{X}} = \vc{\M{A}\M{X}\M{B}'}.$$
\end{prop}

\begin{cor}
  \label{cor:krontimesvec}
  $$(\M{B}' \krn \M{I}) \vc{\M{A}} = \vc{\M{A}\M{B}} = (\M{I} \krn \M{A})\vc{\M{B}}$$
\end{cor}

Note that the size of the identity matrices may not be the same on the left and right sides of the equation, depending on the size of $\M{A}$ and $\M{B}$.

\begin{prop}[{\cite[Prop.~A.18]{TensorTextbook}}]
\label{prop:mixedkron}
For $\M{A}\in\R^{m \times n}$, $\M{B}\in\R^{q \times r}$, $\M{C}\in\R^{n \times p}$, and $\M{D}\in\R^{r \times s}$, we have
$$(\M{A} \krn \M{B})(\M{C} \krn \M{D}) = (\M{A}\M{C} \krn \M{B}\M{D}).$$
\end{prop}

\begin{prop}[{\cite[Prop.~A.23]{TensorTextbook}}]
\label{prop:mkr}
For $\M{A}\in\R^{m \times r}$, $\M{B}\in\R^{n \times r}$, $\M{C}\in\R^{m \times s}$, and $\M{D}\in\R^{n \times s}$, we have
$$(\M{A} \odot \M{B})^{\Tr}(\M{C} \odot \M{D}) = \M{A}'\M{C} \had \M{B}'\M{D}.$$
\end{prop}

\begin{definition}
\label{def:blockdiag}
Let $\Tn{X} \in \R^{m \times n \times p}$ be a 3-way tensor. Define the block diagonal operator $\texttt{blkdiag}_{1}(\Tn{X})$ to take the horizontal slices of $\Tn{X}$ and arrange them as the diagonal entries of a $m\times m$ block matrix such that:

$$
\normalfont{\texttt{blkdiag}}_{1}(\Tn{X}) =
\begin{bmatrix}
  \Tn{X}(1,:,:) & & \M{0}\\
  & \ddots & \\
  \M{0} & & \Tn{X}(m,:,:)
\end{bmatrix} 
= \sum_{i=1}^{m} \M{E}_{ii} \krn \Tn{X}(i,:,:),
$$
where $\M{E}_{ii}=\Vc{e}{i}\Vc{e}{i}'$ is an indicator matrix.
We can similarly define block diagonalization using lateral and frontal slices. The subscript indicates which mode of the tensor remains fixed for each diagonal block sub-matrix.

\end{definition}

\subsection{CP Problem} \label{sec:cpproblem}

In the standard CP approximation problem, we seek factor matrices of a CP model that can be used to approximate the input tensor as accurately as possible in a least squares sense. 

\begin{definition}
\label{def:cp_opt}
Letting $\Tn{X} \in \R^{m\times n \times p}$ be a 3-way tensor,  the CP optimization problem is defined as 
\begin{equation*}
\min_{\M{A}, \M{B} ,\M{C} } \hspace{2px} f(\M{A},\M{B},\M{C}) \equiv \frac{1}{2}\left\| [\![\M{A},\M{B},\M{C}]\!] - \Tn{X}\right\|^2 \text{ subject to } \M{A} \in \R^{m\times r},\M{B} \in \R^{n\times r},\M{C} \in \R^{p\times r},
\end{equation*}
or equivalently, to match the notation of \cref{def:leastsquarederivatives},
\begin{equation*}
\min_{\vv} f(\vv) \equiv \frac12 \left\|\Vc{r}(\vv)\right\|^2 \text{\quad with \quad} \vv = \begin{bmatrix} \vc{\M{A}} \\ \vc{\M{B}} \\ \vc{\M{C}} \end{bmatrix} \in \R^{mr+nr+pr} \text{\quad and \quad} \Vc{r}(\vv) = \vc{[\![\M{A},\M{B},\M{C}]\!] - \Tn{X}}.
\end{equation*}
\end{definition}
Thus, this CP approximation problem is a nonlinear least squares problem with the following Jacobian and gradient.

\begin{lemma}{\cite{ADK11,TensorTextbook}} 
\label{lem:cpgradient}
For $f(\vv) = \frac12 \left\|\Vc{r}(\vv)\right\|^2$ as given by \cref{def:cp_opt}, the Jacobian of $\Vc{r}$ is 
$$\M{J}(\vv) = \begin{bmatrix} \FD{\Vc{r}}{\vc{\A}} & \FD{\Vc{r}}{\vc{\B}} & \FD{\Vc{r}}{\vc{\C}} \end{bmatrix} = \begin{bmatrix}  \left( \M{C} \odot \M{B} \right) \otimes \M{I}_{m} & \M{P}{2} \left[\left ( \M{C} \odot \M{A} \right ) \otimes \M{I}_{n}\right] & \M{P}{3} \left[\left ( \M{B} \odot \M{A} \right ) \otimes \M{I}_{p}\right] \end{bmatrix},$$
where $\M{P}{2}$ and $\M{P}{3}$ are tensor perfect shuffle permutations, and the gradient of $f$ is
$$\Vc{g}\left ( \Vc{v} \right ) = \M{J}^{T}(\vv) \cdot \Vc{r}(\vv) = \begin{bmatrix} \vc{\FD{f}{\A}} \\ \vc{\FD{f}{\B}} \\ \vc{\FD{f}{\C}} \end{bmatrix} 
= \begin{bmatrix} \vc{\A \prn{ \C'\C" \had \B'\B"} - \Tm{X}{1} (\C \krp \B)} \\ \vc{\B \prn{ \C'\C" \had \A'\A" } - \Tm{X}{2} (\C \krp \A)} \\ \vc{\C \prn{ \B'\B" \had \A'\A" } - \Tm{X}{3} (\B \krp \A)} \end{bmatrix}.$$
\end{lemma}

The Alternating Least Squares (ALS) algorithm for CP \cite{CC70,Harshman70} alternates updates to each factor matrix with all others fixed.
In terms of the gradient, it alternates solving $\FD{f}{\A}=\M{0}$ for $\A$, $\FD{f}{\B}=\M{0}$ for $\B$, and $\FD{f}{\C}=\M{0}$ for $\C$, which are all linear systems (with multiple right hand sides). 
By contrast, gradient-based optimization methods, including the Gauss-Newton method below, use the gradient to try to optimize all factor matrices at once at each iteration.

\paragraph{Complexity Analysis}

The cost of computing CP gradient is dominated by the computations involving the data tensor.
The expressions $\Tm{X}{1} (\C \krp \B)$, $\Tm{X}{2} (\C \krp \A)$, and $\Tm{X}{3} (\B \krp \A)$ are called Matricized-Tensor Times Khatri-Rao Products (MTTKRPs).
The computational cost of evaluating the MTTKRPs in all modes is $4mnpr$ plus lower order terms.
The cost of evaluating the other expressions of the gradient are $O((m+n+p)r^2)$, which is also dominated by that of the MTTKRPs.
Note that the cost of an ALS iteration (updating all factor matrices once) is the same, up to the lower order terms.
For more details, see \cite[Sec.~3.6.2]{TensorTextbook}.

\subsection{Gauss-Newton Method for CP}
\label{sec:GN}

The Gauss-Newton method aims to find an improving correction step $\vs_k^{GN}$ from $\vv_k$ by minimizing the
first-order approximation $m_k(\vs) \equiv \frac{1}{2}\|\vr(\vv_k)+\MJ(\vv_k)\vs\|^2$ to $f(\vv_k+\vs)$.
The solution $\vs_k^{GN}$ to this problem satisfies the Gauss-Newton linear system
\begin{equation}
\label{GN}
\MJ^{\Tr}(\vv_k) \MJ(\vv_k) \vs_k^{GN} = - \MJ^{\Tr}(\vv_k) \V{r}(\vv_k) = -\V{g}(\vv_k),
\end{equation}
but when $n$ is large, it is usual to approximate the solution using
an iterative (normally, Krylov) method that requires repeated matrix-vector 
products $\MJ^{\Tr}(\vv_k) \MJ(\vv_k) \Vc[\bar]{v}$ for a sequence of vectors $\Vc[\bar]{v}$.
For example, Vervliet and De Lathauwer \cite{VL19} use preconditioned conjugate gradients with a block-diagonal preconditioner.
Tensorlab \cite{vervlietTensorlab30}
implements three Gauss-Newton variants. The first, {\tt nls\_gndl},
uses a dogleg trust region strategy, in which the minimizer of $m_k$ along 
a piecewise linear path connecting the initial point $\vv_k$ to the 
Gauss-Newton point $\vv_k + \vs_k^{GN}$
via the intermediary Cauchy point $\vv_k + \vs_k^{C}$, where
$\vs_k^{C} = \arg\min m_k(-\alpha\vg(\vv_k))$, is sought within the trust-region
$\{ \vs : \|\vs\| \leq \Delta_k\}$ for a given radius $\Delta_k>0$. The second,
{\tt nls\_gncgs}, applies the conjugate-gradient method to generate 
approximations to the minimizer of $m_k$, but stops whenever one of these
approximations reaches the trust-region boundary. The last, {\tt nls\_lm},
is a Levenberg-Morrison-Marquardt regularization method of which we shall say more in 
\cref{sec:globalization}.

For the CP approximation problem, as outlined in \cref{lem:cpgradient}, we partition $\M{J}(\vv)$ into three column blocks corresponding to the factor matrices.
Then $\M{J}'(\vv)\M{J}(\vv)$ has $3\times 3$ block matrix structure with blocks corresponding to pairs of factor matrices. 
Likewise, we partition the vector $\Vc[\bar]{v}$ to match the factor matrices:
$$
\Vc{\bar{v}} = \begin{bmatrix}
  \vc{\M{\bar{A}}} \\ 
  \vc{\M{\bar{B}}} \\ 
  \vc{\M{\bar{C}}} \\ 
\end{bmatrix}. 
$$
We take advantage of the structure of $\M{J}(\vv)$ to achieve efficient computation of $\MJ^{\Tr}(\vv_k) \MJ(\vv_k) \Vc[\bar]{v}$. 
As shown in \cite{TensorTextbook,VL19}, the computations can be performed as follows:
\begin{equation}
\label{eq:applyJtJ}
\M{J}'(\vv)\M{J}(\vv)\Vc{\bar{v}} =
\begin{bmatrix}
  \vc{\M{\bar{A}}\big(\M{B}'\M{B}" \had \M{C}'\M{C}"\big)
    + \M{A}\big(\M{\bar{B}}'\M{B}" \had \M{C}'\M{C}"\big)
    + \M{A}\big(\M{B}'\M{B}" \had \M{\bar{C}}'\M{C}"\big)} \\[0.5em]
  \vc{\M{B}\big(\M{\bar{A}}'\M{A}" \had \M{C}'\M{C}"\big)
    + \M{\bar{B}}\big(\M{A}'\M{A}" \had \M{C}'\M{C}"\big)
    + \M{B}\big(\M{A}'\M{A}" \had \M{\bar{C}}'\M{C}"\big)} \\[0.5em]
  \vc{\M{C}\big(\M{\bar{A}}'\M{A}" \had \M{B}'\M{B}"\big)
    + \M{C}\big(\M{A}'\M{A}" \had \M{\bar{B}}'\M{B}"\big)
    + \M{\bar{C}}\big(\M{A}'\M{A}" \had \M{B}'\M{B}"\big)}
\end{bmatrix}.
\end{equation}

Note that this formula does not depend on the data tensor and instead requires only matrix multiplications among matrices the sizes of the factor matrices.
Also, matrices $\A'\A$, $\B'\B$, and $\C'\C$ are independent of $\Vc[\bar]{\vv}$.
They are used in evaluating the gradient (see \cref{lem:cpgradient}) and appear multiple times in \cref{eq:applyJtJ}, so they are precomputed once per Gauss-Newton iteration and not recomputed every time $\M{J}'(\vv)\M{J}(\vv)$ is applied to a vector.

\paragraph{Complexity Analysis}
The cost of applying $\M{J}'(\vv)\M{J}(\vv)$ to a vector is small compared to computing the gradient. 
Evaluating \cref{eq:applyJtJ} costs $O((m+n+p)r^2)$, while the gradient computation is $O(mnpr)$ as described in \cref{sec:cpproblem}.
Applying the block-diagonal preconditioner has comparable cost to evaluating \cref{eq:applyJtJ} as well.
Thus, as long as the number of inner iterations in the solve for $\Vc{s}{k}$ is not excessive, the cost of a Gauss-Newton iteration is roughly that of evaluating the gradient.
Note that the precomputation of $\A'\A$, $\B'\B$, and $\C'\C$ saves only a constant factor per inner iteration.
See chapter 13 of \cite{TensorTextbook} for the complete derivation and complexity analysis.

\section{Newton's Method}
\label{sec:algorithm}
We now present our proposed method for computing a CP decomposition via Newton's method.
We will follow the approach of efficient Gauss-Newton algorithms for CP \cite{VL19}, using an iterative method to compute the search direction with an efficient means of applying the (now true) Hessian to a vector that exploits the tensor structure.
While the linear solve for the Gauss-Newton step direction involves a symmetric positive semi-definite matrix, for which iterative approximations using CGLS or LSQR \cite{PaigeSaunders1982} are appropriate,
the true Hessian may be indefinite, so it might appear that we need to use a more sophisticated (and
expensive) iterative linear solver such as SYMMLQ \cite{PaigeSaunders1975}
in this case. Fortunately this is not so.
To enhance the robustness of Newton's method, we will employ different globalization and regularization techniques from those used for Gauss-Newton.

\subsection{Globalization and Regularization} 
\label{sec:globalization}

The CP problem introduced in Section~\ref{sec:cpproblem} is a specific example
of the ubiquitous nonlinear least-squares problem
$$\min_{v \in \R^{r(m+n+p)}} \hspace{2px} f(\vv) \equiv \frac{1}{2}\|\vr(\vv)\|^2$$
in the special case where the unknowns are organized as $\vv = \begin{bmatrix} \vc{\M{A}} \\ \vc{\M{B}} \\ \vc{\M{C}} \end{bmatrix}$ and the $mnp$
residuals are given by $\vr(\vv) = \vc{[\![\M{A},\M{B},\M{C}]\!] - \Tn{X}}$. In this section, we 
focus on this general problem involving $f(\vv)$, and in particular on methods
that are guaranteed (at worst) to find a stationary point of $f$
starting from an arbitrary initial guess;
in most cases this will also be a local minimizer. As we have already seen,
local optimization algorithms, such as the Gauss-Newton and Newton methods, 
traditionally make use of the gradient
$$\vg(\vv) \equiv \MJ^{\Tr}(\vv) \vr(\vv)$$
and often the Hessian
$$\MH(\vv) \equiv \MJ^{\Tr}(\vv) \MJ(\vv) + \sum_{i=1}^m r_i(\vv) \MH_i(\vv)$$
of $f(\vv)$, or a suitable approximation of the latter; here $\MJ(\vv)$ is the 
Jacobian of $\vr(\vv)$, that is the matrix whose $i$-th row is the transpose of 
the gradient, $g_i(\vv)$, of the $i$-th component, $r_i(\vv)$, of $\vr(\vv)$, 
and $\MH_i(\vv)$ is the Hessian of $r_i(\vv)$. However although these methods 
may attain fast local rates of convergence, they are not guaranteed to 
converge from afar. Our focus thus necessarily shifts to methods that do.

Two important classes of optimization algorithm for generic continuous 
optimization are line-search \cite{DennSchn83,NoceWrig06}
and regularization/trust-region \cite{CartGoulToin22,ConnGoulToin00,NoceWrig06}
methods.
The former move from a candidate solution $\vv_k$ to a better one by stepping
along an $f$-improving line (or arc) based on $\vg(\vv_k)$ and sometimes 
$\MH(\vv_k)$. 
The others model $f(\vv)$ around $\vv_k$ by building a suitable approximation 
$m_k(\vs)$ of $f(\vv_k+\vs)$ using $\vg(\vv_k)$ and sometimes $\MH(\vv_k)$,
pick a step $\vs_k$ from $\vv_k$ to reduce $m_k(\vs)$, and then assess whether
$f(\vv_k + \vs_k)$ is suitably smaller than $f(\vv_k)$.

We shall focus here on regularization methods since they have a good reputation
for nonlinear least-squares problems, although most of what we say applies
equally to trust-region methods. The simplest is the 
Levenberg-Morrison-Marquardt (LMM) \cite{Leve44,Marq63,Morr60} method. 
Here $f(\vv_k+\vs)$ is replaced by the model
$$m_{1,k}(\vs) = \frac{1}{2}\|\vr(\vv_k)+\MJ(\vv_k)\vs\|^2 
 + \frac{1}{2}\lambda_k\|\vs\|^2 $$
for which $\vr(\vv_k)+\MJ(\vv_k)\vs$ is the first-order Taylor approximation 
of $\vr(\vv_k+\vs)$, and $\lambda_k > 0$ is an evolving regularization weight that 
implicitly controls the length of the step $\vs$. The step $\vs_k$ is then
chosen to be any $\vs$ for which $m_{1,k}(\vs_k) \leq m_{1,k}(\vs^C_k)$, where 
the Cauchy point $\vs^C_k = \arg\min m_{1,k}(-\alpha \vg(\vv_k))$ gives the 
smallest value of  $m_{1,k}(\vs)$ in the steepest descent direction, 
$-\alpha \vg(\vv_k)$. In particular, $\vs_k$ might be chosen as the minimizer
$\vs_{k,\star}$ of $m_{1,k}(\vs)$, for which
$$ ( \MJ^{\Tr}(\vv_k) \MJ(\vv_k) + \lambda_k I ) \vs_{k,\star} 
 = - \MJ^{\Tr}(\vv_k) \vr(\vv_k).$$
Notice that these equations involve a non-zero diagonal perturbation
to the Gauss-Newton ones \eqref{GN}, and it is this that is crucial
to guarantee global convergence. The rest of the algorithm is straightforward.
Given $\vs_k$, the ratio 
\begin{equation}
\label{rho}
\rho_k = \frac{ f(\vv_k) - f(\vv_k + \vs_k)}{m_{1,k}(0) - m_{1,k}(\vs_k)}
\end{equation}
of the actual to (model) predicted decrease is computed, and so
long as this is adequate (say $\rho_k \geq 0.75$), the correction
$\vv_k + \vs_k$ becomes the new $\vv_{k+1}$ and $\lambda_k$ may be decreased.
By contrast, if the decrease is inadequate, $\vv_{k+1}$ stays at $\vv_k$,
but now $\lambda_k$ is increased; the impact of this is to ensure that
the next step will be smaller than the current one, and thus $m_{k+1}(s)$
resembles $f(\vv_k + \vs)$ better, and ultimately progress will be made.
Although it is possible to use matrix factorization to find the minimizer of
$m_{1,k}(\vs)$, this is likely costly in our case, and instead we use
a Krylov method (GLRT) \cite[Chap. 9 \& 10]{CartGoulToin22}, that simply
involves  $\MJ^{\Tr}(\vv) \MJ(\vv) \Vc[\bar]{v}$ products, to do so.

When the optimal residuals $\vr(\vv_*)$ are small, LMM is an excellent 
method, and fast linear or even superlinear convergence is observed.
However, for problems with larger optimal residuals, the convergence
is frequently observably poorer, and we need to move to something more
Newton like. To do this, we instead consider the regularized Newton
(second-order Taylor) model
\begin{equation}
\label{eq:newton-model}
m_{2,k}(\vs) = f(\vv_k) + \vg'(\vv_k) \vs + 
\frac{1}{2} \vs' \MH(\vv_k) \vs  + \frac{1}{3}\lambda_k\|\vs\|^3
\end{equation}
that includes the $\MK(\vv_k)$ term that is missing in LMM; the
regularization power has to rise above two to compensate for this extra
Hessian term to ensure that the model is bounded from below. All of the
remaining ingredients are as before. We need to find an approximate
minimizer $\vs_k$ of $m_{2,k}(\vs)$ that is at least as good as that at the
Cauchy point $\vs^C_k = \arg\min m_{2,k}(-\alpha \vg(\vv_k))$, we
accept and reject steps and adjust $\lambda_k$  based on the
equivalent of \eqref{rho} involving $m_{2,k}$, and we can find $\vs_k$ 
using either factorization or, in our case,  GLRT involving Hessian-vector 
products.

One important aspect that is peculiar to our tensor fitting problem is that of objective-function ambiguity described in \cref{sec:background:tensor}. 
The scaling ambiguity is unfortunate,
since it implies that there are an infinite number of global minimizers,
and consequently statements about rates of convergence then need to be
phrased in terms of a solution set, not isolated single points. 
Since this is awkward both in theory and practice,
we aim to avoid the issue by insisting that all of
the iterates we generate are normalized so that
$\|\vha\| = \|\vhb\| = \|\vhc\|$.
That is, given $\va, \vb, \vcc$, we \textbf{equilibrate} them via
\begin{equation}
\label{eq:equilibrate}
\va \leftarrow \sigma \frac{\va}{\|\va\|}, \;
\vb \leftarrow \sigma \frac{\vb}{\|\vb\|} \;\text{\quad and \quad}\;
\vcc \leftarrow \sigma \frac{\vcc}{\|\vcc\|}, \;
\;\text{\quad where \quad}\; \sigma = \sqrt[3]{\|\va\| \|\vb\| \|\vcc\|},
\end{equation}
as this gives the required renormalization. 
This is the same equilibration used by Uschmajew for the ALS algorithm \cite{Uschmajew12} and is implemented in Tensor Toolbox's \texttt{normalize} function \cite{TensorToolbox}.
Crucially, this does not
impact the global convergence of the method, as this equilibration is
akin to taking an evaluation-free magical step 
\cite[\S10.4.1]{ConnGoulToin00}, but what the precise
implication for the rate of convergence is an open question---our 
experiments suggest that a fast rate, when expected, is retained.

The other, less serious ambiguity is that we may arbitrarily (but consistently) 
permute the columns of $\A$, $\B$ and $\C$ without changing the generated Kruskal tensor. We 
ignore this possibility in practice as it leads to distinct, as opposed to continuous, solution 
sets.

\subsection{Deriving the Hessian for CP}
\label{sec:algorithm:der_hess}

Next, we derive the true Hessian for a 3-way CP decomposition $\Tn{X} \approx \dsqr{\M{A}, \M{B}, \M{C}} \in \R^{m \times n \times p}$ such that $\M{A} \in \R^{m \times r}$, $\M{B} \in \R^{n \times r}$, and $\M{C} \in \R^{p \times r}$. 
We use the indices $i$, $j$, $k$, and $\ell$ for $m$, $n$, $p$, and $r$, respectively. 

The nonlinear least squares optimization for CP is given in \cref{def:cp_opt}, and as shown in \cref{def:leastsquarederivatives}, the Hessian can be expressed as 
$$\MH(\vv) \equiv \MJ^{\Tr}(\vv) \MJ(\vv) + \MK(\vv)\text{, \quad where \quad } \MK(\vv) = \sum_{i,j,k} r_{ijk}(\vv) \nabla^2 r_{ijk}(\vv) \text{ \quad and \quad } \vv = \begin{bmatrix} \vc{\M{A}} \\ \vc{\M{B}} \\ \vc{\M{C}} \end{bmatrix}.$$
As the structure of $\MJ^{\Tr}(\vv) \MJ(\vv)$ and efficient means of applying it to a vector are described in \cref{sec:GN}, we focus now on $\MK(\vv)$.
Here $\Tn{R}=\dsqr{\M{A}, \M{B}, \M{C}}-\Tn{X}$ is the residual tensor and $\MH_{ijk}(\vv)$ is the Hessian of the scalar-valued function $r_{ijk}(\vv)$, an element of $\Tn{R}$.

\begin{figure}
  \centering
  \includegraphics[width=0.6\textwidth]{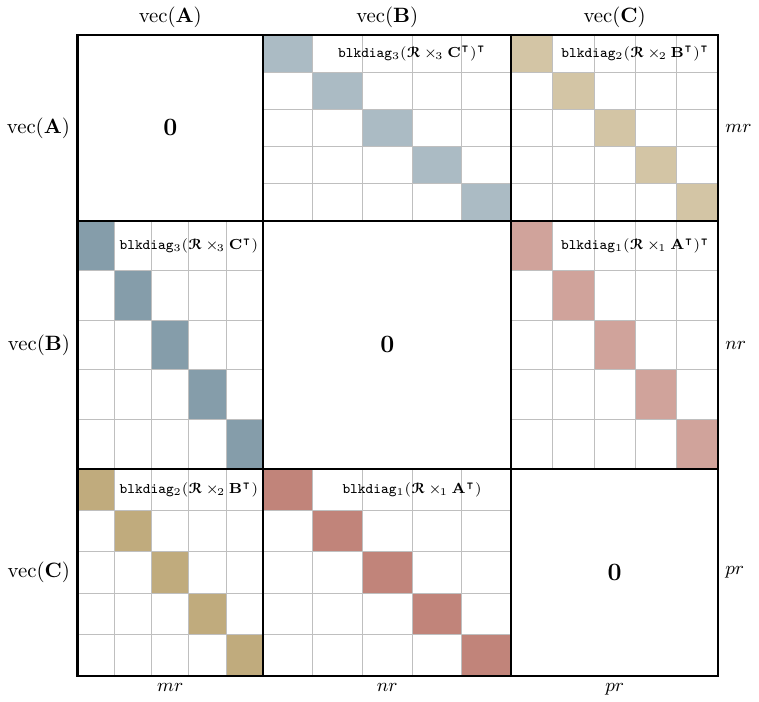}
  \caption{Diagram of $\M{K}$'s structure in the case that the input is an $m\times n\times p$ tensor and $r=5$. }
  \label{fig:hessian-K}
\end{figure}

Similar to the structure of $\MJ^{\Tr}(\vv) \MJ(\vv)$, we consider $\MK$ (dropping the dependence on $\vv$ for ease of notation) to be a symmetric $3\times3$ block matrix based on the mixed partials with respect to each pair of factor matrices:
\begin{equation}
    \label{eq:K}
\M{K} = 
\begin{bmatrix}
  \M{K}{AA}" & \M{K}{AB}" & \M{K}{AC}"\\
  \M{K}{AB}' & \M{K}{BB}" & \M{K}{BC}" \\
  \M{K}{AC}' & \M{K}{BC}' & \M{K}{CC}"
\end{bmatrix}, \text{\quad where, for example, \quad} 
\M{K}{AB} = \sum_{ijk} r_{ijk} \, \SD{r_{ijk}}{\vc{\M{B}}}{\vc{\M{A}}}.
\end{equation}
Note that $r_{ijk}(\vv) = \sum_{\ell=1}^r a_{i\ell} b_{j\ell} c_{k\ell} - x_{ijk}$, which is a linear function in each of the factor matrices.
This implies that any mixed partial derivative involving two elements of the same factor matrix is 0, or equivalently, $\M{K}{AA}=\M{0}$, $\M{K}{BB}=\M{0}$, and $\M{K}{CC}=\M{0}$.
We now derive expressions for the three off-diagonal blocks.
A visualization of the sparsity structure of $\M{K}$ is given in \cref{fig:hessian-K}.

\begin{lemma}
\label{lemma:newton3dmixedpartial}
The off-diagonal blocks of the matrix $\M{K}$ as defined in \cref{eq:K} are given by
  $$\M{K}{AB} = \normalfont{\texttt{blkdiag}}_{3} \left(\Tn{R} \times_3 \M{C}'\right)$$
  $$\M{K}{AC} = \normalfont{\texttt{blkdiag}}_{2} \left(\Tn{R} \times_2 \M{B}'\right)$$
  $$\M{K}{BC} = \normalfont{\texttt{blkdiag}}_{1} \left(\Tn{R} \times_1 \M{A}'\right)$$
  with ${\normalfont \texttt{blkdiag}}_k\left(\cdot\right)$ given by \cref{def:blockdiag} and $\Tn{R}=\dsqr{\M{A}, \M{B}, \M{C}}-\Tn{X}$.
\end{lemma}

\begin{proof}
By definition, $\M{K}{AB} = \sum_{ijk} r_{ijk} \frac{\partial^2 r_{ijk}}{\partial \vc{\M{B}}\, \partial \vc{\M{A}}}$. 
We will first find the mixed partial of a single residual element, then we will sum over all residual elements.

First, we obtain the gradient of a single residual element $r_{ijk}$ with respect to $\vc{\M{A}}$. 
Let $\Tn{E}{ijk}$ be the 3-way indicator tensor, which is $\M{E}{i, \mathbb{L}(j,k)}$ when unfolded on the first mode.
The Jacobian of $\Vc{r}$ is given in \cref{lem:cpgradient}, and in particular, $\FD{\Vc{r}}{\vc{\A}} = \left(\M{C} \krp \M{B}\right) \krn \M{I}{m}$.
Thus, we can find the gradient of $r_{ijk}$ (with respect to $\vc{\A}$) by selecting and transposing the corresponding row of the Jacobian of $\Vc{r}$ (with respect to $\vc{\A}$):
\begin{align*}
\frac{\partial r_{ijk}}{\partial \vc{\M{A}}} 
&= \left(\vc{\Tn{E}{ijk}}^{\Tr} \left[\left(\M{C} \krp \M{B}\right) \krn \M{I}{m}\right]\right)^{\Tr} \\
&= \left[\left(\M{C} \krp \M{B}\right)^{\Tr} \krn \M{I}{m}\right] \vc{\M{E}{i, \mathbb{L}(j,k)}} & \text{Unfold } \Tn{E}{ijk} \text{ on the first mode}\\
&= \left(\M{I}{r} \krn \M{E}{i, \mathbb{L}(j,k)}\right) \vc{\M{C} \krp \M{B}} & \cref{cor:krontimesvec}.
\end{align*}
  
Next, we differentiate with respect to $\vc{\B}$.
Since $\M{I}{r} \krn \M{E}{i, \mathbb{L}(j,k)}$ does not depend on $\M{B}$, we differentiate $\vc{\M{C} \krp \M{B}}$ with respect to $\vc{\M{B}}$ and multiply back. 
By definition, the $\ell$-th column of $\M{C} \krp \M{B}$ is $\Vc{c}{\ell} \krn \Vc{b}{\ell}$, so
\begin{align*}
\vc{\M{C} \krp \M{B}} &= \sum_{\ell=1}^{r} \Vc{e}{\ell} \krn \Vc{c}{\ell} \krn \Vc{b}{\ell} 
= \sum_{\ell=1}^{r} \left(\Vc{e}{\ell} \krn \Vc{c}{\ell} \krn \M{I}_n\right) \Vc{b}{\ell}  & \cref{prop:mixedkron} \\
&= \sum_{\ell=1}^{r} \left(\Vc{e}{\ell} \krn \Vc{c}{\ell} \krn \M{I}_n\right) \M{B}\Vc{e}{\ell} 
= \sum_{\ell=1}^{r} \left(\Vc{e}{\ell} \krn \Vc{c}{\ell} \krn \M{I}_n\right) \left(\Vc{e}{\ell}' \krn \M{I}{n}\right)\vc{\M{B}} & \cref{cor:krontimesvec}. \\
\end{align*}
Thus,
$$\FD{\vc{\M{C} \krp \M{B}}}{\vc{\B}} = \sum_{\ell=1}^{r}  \left(\Vc{e}{\ell} \krn \Vc{c}{\ell} \krn \M{I}_n\right) \left(\Vc{e}{\ell}' \krn \M{I}{n}\right),$$
and therefore
  $$\frac{\partial^2 r_{ijk}}{\partial \vc{\M{B}}\, \partial \vc{\M{A}}} = \left(\M{I}{r} \krn \M{E}{i, \mathbb{L}(j,k)}\right) \sum_{\ell=1}^r \left(\Vc{e}{\ell} \krn \Vc{c}{\ell} \krn \M{I}{n}\right)\!\left(\Vc{e}{\ell}' \krn \M{I}{n}\right).$$
  We simplify the product $\left(\M{I}{r} \krn \M{E}{i, \mathbb{L}(j,k)}\right)\!\left(\Vc{e}{\ell} \krn \Vc{c}{\ell} \krn \M{I}{n}\right)$ using \cref{prop:mixedkron}:
  \begin{align*}
    \left[\M{I}{r} \krn \M{E}{i, \mathbb{L}(j,k)}\right]\!\left(\Vc{e}{\ell} \krn \left(\Vc{c}{\ell} \krn \M{I}{n}\right)\right)
    &= \Vc{e}{\ell} \krn \left[\M{E}{i, \mathbb{L}(j,k)}\!\left(\Vc{c}{\ell} \krn \M{I}{n}\right)\right].
  \end{align*}
  To evaluate the right factor, recall $\M{E}{i, \mathbb{L}(j,k)} = \Vc{e}{i}\left(\Vc{e}{k} \krn \Vc{e}{j}\right)^{\Tr}$ and apply \cref{prop:mixedkron}:
  $$\M{E}{i, \mathbb{L}(j,k)}\!\left(\Vc{c}{\ell} \krn \M{I}{n}\right) = \Vc{e}{i}\left(\Vc{e}{k}'\Vc{c}{\ell} \krn \Vc{e}{j}'\right) = c_{k\ell}\M{E}{ij}.$$
  Substituting back and applying \cref{prop:mixedkron} once more:
  $$\frac{\partial^2 r_{ijk}}{\partial \vc{\M{B}}\, \partial \vc{\M{A}}} = \sum_{\ell=1}^r \left(\Vc{e}{\ell} \krn c_{k\ell}\M{E}{ij}\right)\!\left(\Vc{e}{\ell}' \krn \M{I}{n}\right) = \sum_{\ell=1}^r \Vc{e}{\ell}\Vc{e}{\ell}' \krn \left(c_{k\ell}\M{E}{ij}\right).$$
  Multiplying by $r_{ijk}$, summing over $(i,j,k)$, and exchanging the order of summation, we have
  \begin{align*}
    \M{K}{AB}
    &= \sum_{\ell=1}^r \Vc{e}{\ell}\Vc{e}{\ell}' \krn \sum_{ijk} r_{ijk}\, c_{k\ell}\, \M{E}{ij} \\
    &= \sum_{\ell=1}^r \M{E}_{\ell\ell} \krn \left(\Tn{R} \ttm[3] \Vc{c}{\ell}'\right) \\
    &= \texttt{blkdiag}_{3}\!\left(\Tn{R} \ttm[3] \M{C}'\right) \qquad \cref{def:blockdiag}.
  \end{align*}
The derivations of $\M{K}{AC}$ and $\M{K}{BC}$ are analogous.
\end{proof}

\subsection{Efficiently Applying the Hessian to a Vector}
\label{sec:algorithm:apply}

To apply the Hessian $\M{H}(\vv)$ to a vector $\Vc[\bar]{\vv}$, we separately compute $\M{J}'(\vv)\M{J}(\vv)\Vc[\bar]{\vv}$ (as described in \cref{sec:GN}) and $\M{K}(\vv)\Vc[\bar]{\vv}$.
As before, we will break the vector $\Vc{\bar{v}}$ into three components which correspond to the three factor matrices, and we drop the dependence on $\vv$ to ease notation.  
We have
\begin{equation}
\label{eq:Kvbar}
\M{K} \Vc{\bar{v}} = 
\begin{bmatrix}
  \M{0} & \M{K}{AB}" & \M{K}{AC}"\\
  \M{K}{AB}' & \M{0}  & \M{K}{BC}" \\
  \M{K}{AC}' & \M{K}{BC}' & \M{0} 
\end{bmatrix}
\begin{bmatrix}
  \vc{\M{\bar{A}}} \\ \vc{\M{\bar{B}}} \\ \vc{\M{\bar{C}}}
\end{bmatrix}
=
\begin{bmatrix}
  \begin{aligned}{} 
    \M{K}{AB}" \vc{\M{\bar{B}}} &+ \M{K}{AC}" \vc{\M{\bar{C}}} \\
    \M{K}{AB}' \vc{\M{\bar{A}}} &+ \M{K}{BC}" \vc{\M{\bar{C}}} \\
    \M{K}{AC}' \vc{\M{\bar{A}}} &+ \M{K}{BC}' \vc{\M{\bar{B}}}
  \end{aligned}
\end{bmatrix}.
\end{equation}

The following lemma illustrates that the structure of these computations is similar to the gradient, but with mixed contributions from the CP model vector $\vv$ and the arbitrary vector $\Vc[\bar]{v}$.

\begin{lemma}
\label{lem:Kvbar}
The components of $\M{K}\Vc[\bar]{v}$ from \cref{eq:Kvbar} can be computed as follows: 
\begin{align*}
\M{K}{AB}\vc{\M[\bar]{B}} &= \vc{\Tm{R}{1}\left(\M{C}\krp \M{\bar{B}}\right)} = \vc{\A(\C'\C \had \B'\M[\bar]{B}) - \Tm{X}{1}\left(\M{C}\krp \M{\bar{B}}\right)}\\
\M{K}{AC}\vc{\M[\bar]{C}} &= \vc{\Tm{R}{1}\left(\M[\bar]{C}\krp \M{B}\right)} = \vc{\A(\C'\M[\bar]{C} \had \B'\B) - \Tm{X}{1}\left(\M[\bar]{C}\krp \M{B}\right)} \\
\M{K}{BC}\vc{\M[\bar]{C}} &= \vc{\Tm{R}{2}\left(\M[\bar]{C}\krp \M{A}\right)} = \vc{\B(\C'\M[\bar]{C} \had \A'\A) - \Tm{X}{2}\left(\M[\bar]{C}\krp \M{A}\right)} \\
\M{K}{AB}'\vc{\M[\bar]{A}} &= \vc{\Tm{R}{2}\left(\M{C}\krp \M{\bar{A}}\right)} = \vc{\B(\C'\C \had \A'\M[\bar]{A}) - \Tm{X}{2}\left(\M{C}\krp \M{\bar{A}}\right)} \\
\M{K}{AC}'\vc{\M[\bar]{A}} &= \vc{\Tm{R}{3}\left(\M{B}\krp \M{\bar{A}}\right)} = \vc{\C(\B'\B \had \A'\M[\bar]{A}) - \Tm{X}{3}\left(\M{B}\krp \M{\bar{A}}\right)}\\
\M{K}{BC}'\vc{\M[\bar]{B}} &= \vc{\Tm{R}{3}\left(\M[\bar]{B}\krp \M{A}\right)} = \vc{\C(\B'\M[\bar]{B} \had \A'\A) - \Tm{X}{3}\left(\M[\bar]{B}\krp \M{A}\right)}
\end{align*}

\end{lemma}

\begin{proof}
We detail the derivation only for $\M{K}{AB} \vc{\M{\bar{B}}}=\texttt{blkdiag}_{3}\left( \Tn{R} \ttm[3] \C' \right) \vc{\M{\bar{B}}}$, as the other cases are derived similarly.
Note that by the definition of the mode-$3$ product (see \cref{def:ttm}), the $\ell$th frontal slice of $\Tn{R}\times_3 \C'$ is $\Tn{R} \ttm[3] \Vc{c}{\ell}' = \Tm{R}{1}\left(\Vc{c}{\ell} \krn \M{I}{n}\right)$.
Then
\begin{align*}
  \texttt{blkdiag}_{3}\left( \Tn{R} \ttm[3] \C' \right) \vc{\M{\bar{B}}} &= 
  \left[\sum_{\ell=1}^{r} \M{E}{\ell\ell} \krn (\Tn{R} \ttm[3] \Vc{c}{\ell}') \right] \vc{\M{\bar{B}}} \\
  &= \sum_{\ell=1}^{r} \vc{(\Tn{R} \ttm[3] \Vc{c}{\ell}')\M{\bar{B}} \Vc{e}{\ell}"\Vc{e}{\ell}'} & \cref{prop:krontimesvec} \\
  &= \sum_{\ell=1}^{r}\vc{\Tm{R}{1}\left(\Vc{c}{\ell} \krn \M{I}{n}\right)\Vc{\bar{b}}{\ell}" \Vc{e}{\ell}'} \\
  &= \sum_{\ell=1}^{r}\vc{\Tm{R}{1}\left(\Vc{c}{\ell} \krn \Vc{\bar{b}}{\ell}\right) \Vc{e}{\ell}'} & \cref{prop:mixedkron}\\
  &= \sum_{\ell=1}^{r}\vc{\Tm{R}{1}\left(\M{C} \krp \M{\bar{B}}\right)\Vc{e}{\ell}"\Vc{e}{\ell}'} & \cref{def:kr}\\
  &= \vc{\Tm{R}{1} \left(\M{C} \krp \M{\bar{B}}\right)} & \text{ because } \sum_{\ell=1}^{r}\Vc{e}{\ell}"\Vc{e}{\ell}' = \M{I}{r}.
\end{align*}
As $\Tm{R}{1} = \A(\C \krp \B)^{\Tr} - \Tm{X}{1}$, 
$$\Tm{R}{1}\left(\M{C}\krp \M{\bar{B}}\right) = \A(\C'\C \had \B'\M[\bar]{B}) - \Tm{X}{1}\left(\M{C}\krp \M{\bar{B}}\right)$$ 
follows from \cref{prop:mkr}.

\end{proof}

Note the similarities of these expressions to the CP gradient given by \cref{lem:cpgradient}.
In particular, unlike the computation of $\M{J}'\M{J}\Vc[\bar]{v}$ in \cref{eq:applyJtJ}, $\M{K}\Vc[\bar]{v}$ involves the data tensor $\Tn{X}$ in the form of MTTKRPs.
Evaluating these MTTKRPs directly would make every iteration of an iterative solve for $\Vc{s}{k}$ as expensive as evaluating the gradient (roughly equivalent to one Gauss-Newton iteration, for example). 

However, we can avoid this overhead using precomputation of contractions between the data tensor and the Newton iterate.
Every MTTKRP in \cref{lem:Kvbar} involves one factor matrix from $\Vc{v}$ (the fixed Newton iterate) and one factor matrix from $\Vc[\bar]{v}$ (the arbitrary vector that varies over iterations).
For example, $\M{K}{AB}'\vc{\M[\bar]{A}}$ involves the MTTKRP $\M{M}=\Tm{X}{2}\left(\M{C}\krp \M{\bar{A}}\right)$.
As described in \cite[Sec.~3.5.1]{TensorTextbook}, we can compute the MTTKRP using the following two steps: (1) $\Tn{Y}{AB}=\Tn{X} \ttm[3] \M{C}'$ and (2) $\vc{\M{M}} = \texttt{blkdiag}_{3}\!\left(\Tn{Y}{AB}\right)^{\Tr}\vc{\M[\bar]{A}}$, which is computed as batched matrix-vector products.
In fact, the six terms can all be computed from three precomputations involving $\Tn{X}$ and $\vv$: $\Tn{Y}{BC}=\Tn{X} \ttm[1] \A'$, $\Tn{Y}{AC}=\Tn{X} \ttm[2] \B'$, and $\Tn{Y}{AB}=\Tn{X} \ttm[3] \C'$.
Because of the overlap, we can also use these intermediate quantities (as well as $\A'\A$, $\B'\B$, and $\C'\C$) to evaluate the gradient.

\paragraph{Complexity Analysis}
The cost of applying $\M{H}(\vv)$ to a vector is split into costs per Newton iteration and cost per GLRT iteration.
The precomputations in \cref{alg:precomputation}, which are performed every Newton iteration, include the three TTMs with the data tensor and computing the Gram matrix for each factor matrix, and they have computational cost $6mnpr$ plus lower order terms.
Note that the gradient can be computed very cheaply given these precomputations (see \cref{alg:comp_grad}), so this cost is in place of rather than in addition to that of the gradient.
That is, the precomputations of a Newton iteration cost roughly 50\% more than computing only the gradient.

The cost per GLRT iteration includes the cost of applying $\M{J}'(\vv)\M{J}(\vv)$ and completing the application of $\M{K}(\vv)$.
The cost of applying $\M{J}'(\vv)\M{J}(\vv)$ is $O((m+n+p)r^2)$ (see \cref{sec:GN}).
The six batched matrix-vector products to complete the application of $\M{K}(\vv)$ have cost $O((mn+np+mp)r)$, and the other operations cost $O((m+n+p)r^2)$.
Therefore, assuming $r$ is small relative to the tensor dimensions, the cost of a GLRT iteration is more expensive when applying the true Hessian as opposed to the Gauss-Newton approximation, but in either case we expect to be dominated by the evaluation of the gradient (i.e., precomputations in the case of Newton's method) at each outer iteration.

\subsection{Algorithm Outline}
\label{sec:algorithm:outline}

The pseudocode for the proposed algorithm is given by \cref{alg:newton_GLRT}, and the key computational subroutines are given in \cref{alg:precomputation,alg:comp_grad,alg:apply_hessian}.
As described in \cref{sec:globalization}, the overall approach involves outer and inner iterations: the outer iteration corresponds to a Newton step and an approximate CP model represented by $\Vc{v}{k}$, and the inner iteration corresponds to a step of the Krylov method GLRT used to compute the update $\Vc{s}{k}$.
We use a stopping criterion for the Newton iteration based on a gradient norm tolerance (see line \ref{line:stop}), and we use the Eisenstat-Walker method to adaptively change the forcing tolerance for GLRT at each Newton iteration \cite{EW96}, though that is not specified in \cref{alg:newton_GLRT}.
Note that we equilibrate the CP model at every Newton iteration as described in \cref{eq:equilibrate} (see line \ref{line:equilibrate}).

The dominant computation of using Newton's method to compute CP can be divided into two major components: evaluation of the gradient and computing the search direction $\vs_k$. The latter involves multiple iterations of the GLRT algorithm, whose cost is dominated by applying the $\M{K}$ term of the Hessian to a vector as described in \cref{sec:algorithm:apply}. 
Both of these major operations can be cast as multi-step tensor contractions. 
For each Newton iteration, we perform the precomputations that contribute to both evaluations as shown in \cref{alg:precomputation}, and then complete the above operations using the result.

\begin{algorithm}
\caption{Precomputations for Gradient and $\M{K}$}
\label{alg:precomputation}
\begin{algorithmic}[1]
    \Function{$\{\Tn{Y}\} = $Precompute}{$\Tn{X}, \Vc{v}$}
    \State Unpack $\Vc{v}$ into $\M{A}, \M{B}, \M{C}$
    \vspace{1.5pt}
	  \State $\M{S}_{A} \gets \M{A}'\M{A}$, \quad $\M{S}_{B} \gets \M{B}'\M{B}$, \quad $\M{S}_{C} \gets \M{C}'\M{C}$
    \vspace{1.5pt}
    \State $\Tn{Y}{AB} \gets \Tn{X} \ttm[3] \M{C}'$
    \State $\Tn{Y}{AC} \gets \Tn{X} \ttm[2] \M{B}'$
    \State $\Tn{Y}{BC} \gets \Tn{X} \ttm[1] \M{A}'$
    \State \Return $\{ \Tn{Y}{AB}, \Tn{Y}{AC}, \Tn{Y}{BC} \}, \{\M{S}_{A}, \M{S}_{B}, \M{S}_{C} \}$
    \EndFunction
    \end{algorithmic}
\end{algorithm}

For the gradient computation, each MTTKRP is completed by one batched matrix-vector product involving a TTM intermediate from $\{\Tn{Y}\}$ that has all modes except two contracted (see \cref{alg:comp_grad}). 
To apply $\M{K}$, for each nonzero block we apply one input-vector factor matrix to an intermediate from $\{\Tn{Y}\}$ as shown in \cref{sec:algorithm:apply}.
The pseudocode is given in \cref{alg:apply_hessian}.
Therefore, by precomputing the $\{\Tn{Y}\}$ intermediates once per Newton step, we avoid redundant computation across the gradient and all $\M{K}$ applications.

\begin{algorithm}
\caption{Completion of Gradient Evaluation}
\label{alg:comp_grad}
\begin{algorithmic}[1]
	\Function{$\vg= $ comp-gradient}{$\Vc{v}, \{\Tn{Y}\}, \{\M{S}\}$} 
    \State Unpack $\Vc{v}$ into $\M{A}, \M{B}, \M{C}$    
    \State $\vc{\M{U}{A}} \gets \texttt{blkdiag}_{3}(\Tn{Y}{AB})^{\phantom \Tr}\vc{\M{B}}$
    \State $\vc{\M{U}{B}} \gets \texttt{blkdiag}_{1}(\Tn{Y}{BC})^{\phantom \Tr}\vc{\M{C}}$ 
    \State $\vc{\M{U}{C}} \gets \texttt{blkdiag}_{2}(\Tn{Y}{AC})^{\Tr}\vc{\M{A}}$
    
	\State $\M{G}{A} \gets \M{A}(\M{S}{B} \had \M{S}{C}) - \M{U}{A}$ 
        \Comment{$\M{G}{A} = \FD{f}{\M{A}}$}
	\State $\M{G}{B} \gets \M{B}(\M{S}{A} \had \M{S}{C}) - \M{U}{B}$
        \Comment{$\M{G}{B} = \FD{f}{\M{B}}$}
	\State $\M{G}{C} \gets \M{C}(\M{S}{A} \had \M{S}{B}) - \M{U}{C}$
        \Comment{$\M{G}{C} = \FD{f}{\M{C}}$}
	\State Pack $\M{G}{A}, \M{G}{B}, \M{G}{C}$ into $\vg$
    \State \Return $\vg$
	\EndFunction
\end{algorithmic}
\end{algorithm}

\begin{algorithm}
\caption{Completion of Application of $\M{K}$ Term of Hessian to a Vector}
\label{alg:apply_hessian}
\begin{algorithmic}[1]
	\Function{$\vs = $Comp-Apply-K}{$\Vc{V}, \Vc{\bar{v}}, \{\Tn{Y}\}, \{\M{S}\}$}
	\State Unpack $\Vc{v}$ into $\M{A}, \M{B}, \M{C}$, unpack $\bar{\Vc{v}}$ into $\bar{\M{A}}, \bar{\M{B}}, \bar{\M{C}}$
    \vspace{1.5pt}
	\State $\M[\bar]{S}_{A} \gets \M{A}'\M[\bar]{A}$, \quad $\M[\bar]{S}_{B} \gets \M{B}'\M[\bar]{B}$, \quad $\M[\bar]{S}_{C} \gets \M{C}'\M[\bar]{C}$
    \vspace{1.5pt}

    \State $\vc{\M{N}{AB}} \gets \texttt{blkdiag}_{3} (\Tn{Y}{AB})^{\phantom \Tr}      \vc{\M[\bar]{B}}$ 
    \State $\vc{\M{N}{AC}} \gets \texttt{blkdiag}_{2} (\Tn{Y}{AC})^{\phantom \Tr}      \vc{\M[\bar]{C}}$
    \State $\vc{\M{N}{BA}} \gets \texttt{blkdiag}_{3} (\Tn{Y}{AB})^{\Tr}\vc{\M[\bar]{A}}$
    \State $\vc{\M{N}{BC}} \gets \texttt{blkdiag}_{1} (\Tn{Y}{BC})^{\phantom \Tr}      \vc{\M[\bar]{C}}$
    \State $\vc{\M{N}{CA}} \gets \texttt{blkdiag}_{2} (\Tn{Y}{AC})^{\Tr}\vc{\M[\bar]{A}}$
    \State $\vc{\M{N}{CB}} \gets \texttt{blkdiag}_{1} (\Tn{Y}{BC})^{\Tr}\vc{\M[\bar]{B}}$
    \vspace{2.5pt}
	\State $\M{Q}{A} \gets \M{A}(\M{S}{C} \had \bar{\M{S}}_{B} + \M{S}{B} \had \bar{\M{S}}_{C}) - \M{N}{AB} - \M{N}{AC}$
    \vspace{2pt}
	\State $\M{Q}{B} \gets \M{B}(\M{S}{C} \had \bar{\M{S}}_{A} + \M{S}{A} \had \bar{\M{S}}_{C}) - \M{N}{BA} - \M{N}{BC}$
    \vspace{2pt}
	\State $\M{Q}{C} \gets \M{C}(\M{S}{B} \had \bar{\M{S}}_{A} + \M{S}{A} \had \bar{\M{S}}_{B}) - \M{N}{CA} - \M{N}{CB}$
    \vspace{2pt}
    \State Pack $\M{Q}{A}, \M{Q}{B}, \M{Q}{C}$ into $\Vc{s}$
	\State \Return $\Vc{s}$
	\EndFunction
\end{algorithmic}
\end{algorithm}

After the search direction $\vs_k$ has been computed, we then determine whether to accept or reject the step using regularization semantics described in \cref{sec:globalization}. We compute the predicted change in the objective using the unregularized Hessian, then compute the ratio $\rho$ between the true and model decrease found from taking step $\vs_k$. If the ratio is sufficiently high, then we accept the step, equilibrate the new point, and decrease $\lambda$. If the ratio is small or negative, then the step is rejected, and $\lambda$ is increased. There are also intermediate values of $\rho$ where the step is accepted, but $\lambda$ does not decrease.

\begin{algorithm}
\caption{CP Newton with GLRT}
\label{alg:newton_GLRT}
\begin{algorithmic}[1]
	\Function{$\vv =$ CP-Newton-GLRT}{$\Tn{X},\dsqr{\M{A}{0},\M{B}{0},\M{C}{0}}, \varepsilon$}
    \State Pack $\M{A}{0}, \M{B}{0}, \M{C}{0}$ into $\vv_0$
    \State Equilibrate $\Vc{v}{0}$ as given by \cref{eq:equilibrate}
	\State $\{\Tn{Y}\}, \{\M{S}\} \gets \Call{precompute}{\Tn{X}, \Vc{v}{0}}$ \Comment{TTM and Gram intermediates, shared by $\vg$ and $\M{K}$}
	\State $\vg_{0} \gets \Call{comp-gradient}{\Vc{v}{0}, \{\Tn{Y}\}, \{\M{S}\}}$
	\State $f_0 \gets \Call{objective}{\Tn{X}, \Vc{v}{0}, \{\Tn{Y}\}, \{\M{S}\}}$
	\For{$k = 1, 2, \dots$}
		\While{GLRT with cubic weight $\lambda$ not converged} \Comment{Compute trial step $\vs_k$}
      \State Apply $\left(\M{J}'\M{J}" + \M{K}\right)\Vc{\bar{v}}$ using $\{\Tn{Y}\}$ and $\{\M{S}\}$ \Comment{for some $\Vc{\bar{v}}$ determined by GLRT} \label{line:apply}
      \State Update $\vs_k$
    \EndWhile
		\State $\Delta_{m} \gets -\left[\tfrac{1}{2}\,\vs_k^{\Tr}\left(\M{J}'\M{J}" + \M{K}\right)\vs_k
        + {\vg_{k-1}}^{\Tr}\vs_k\right]$ \Comment{predicted decrease}
		\State $f_k \gets \Call{objective}{\Tn{X}, \Vc{v}{k-1} + \vs_k}$ \label{line:obj}
		\State $\rho \gets \left(f_{k-1} - f_{k}\right) / \Delta_{m}$
		\If{$\rho > \rho_{accept}$} \Comment{accept the trial step}
			\State $\Vc{v}{k} \gets \Vc{v}{k-1} + \vs_k$
            \State Equilibrate $\Vc{v}{k}$ as given by \cref{eq:equilibrate} \label{line:equilibrate}
        	\State $\{\Tn{Y}\}, \{\M{S}\} \gets \Call{precompute}{\Tn{X}, \Vc{v}{k}}$ \label{line:precompute}
        	\State $\vg_{k} \gets \Call{comp-gradient}{\Vc{v}{k}, \{\Tn{Y}\}, \{\M{S}\}}$ \label{line:comp_grad}
		\Else 
			\State $\Vc{v}{k} \gets \Vc{v}{k-1}$; \quad $f_{k} \gets f_{k-1}$; \quad $\vg_{k} \gets \vg_{k-1}$ \Comment{reject update}
		\EndIf
		\State Adjust $\lambda$ according to $\rho$ \Comment{increase when $\rho$ is small, decrease when large}
		\If{$\left\|\vg_{k}\right\| < \varepsilon$} \label{line:stop}
            \State \Return $\vv_k$ 
        \EndIf
	\EndFor
	\EndFunction
\end{algorithmic}
\end{algorithm}

\paragraph{Complexity Analysis} 

\Cref{tab:complexity} outlines the cost of each of the above operations in outer and inner iterations, assuming $\Tn{X}$ is an $m\times n \times p$ tensor and the CP model has rank $r$.
For each outer iteration, we highlight the calls to the \textsc{precompute} function (line \ref{line:precompute}), which includes computing both TTMs and Gram matrices, and the \textsc{comp-gradient} function (line \ref{line:comp_grad}), which completes the gradient evaluation.
The inner iterations are dominated by the application of the (regularized) Hessian in line \ref{line:apply}, which includes both the application of $\M{J}'\M{J}"$ and $\M{K}$.
We also show the comparison with the complexity of Gauss-Newton methods that rely on a iterative method to compute the step direction.

Note that the cost of a rejected step is a fraction (roughly 1/3rd) of the cost of an accepted step.
While not reflected in the pseudocode of \cref{alg:newton_GLRT} for simplicity, we precompute one of the TTM results (e.g., $\Tn{Y}{AB}$), complete an MTTKRP, and evaluate the objective function value in line~\ref{line:obj}.
If the step is accepted, then the temporary can be equilibrated to match the corresponding factor matrix (rather than recomputed), and the objective function evaluation is essentially free.
If the step is rejected, then the evaluation is an added cost which is less than the cost of the full precomputation subroutine.

\begin{table}
\begin{center}
\begin{tabular}{lcc}
    \toprule
    \textbf{Calculation} & \textbf{Newton Complexity} & \textbf{Gauss-Newton Complexity}  \\
    \midrule
    \multicolumn{2}{l}{\textit{Per Outer Iteration:}} \\[2pt]
    \quad \textbf{Precomputed TTMs} & $6mnpr$ & $4mnpr$\\
    \quad Gradient Completions & $O((mn+mp+np)r)$ & $O((mn+mp+np)r)$ \\
    \quad Gram Matrices & $O((m+n+p)r^2)$ & $O((m+n+p)r^2)$ \\
    \quad Per Rejected Step & $2mnpr$ & $2mnpr$ \\[6pt]
    \multicolumn{2}{l}{\textit{Per Inner Iteration:}} \\[2pt]
    \quad Apply $\M{J}'\M{J}"$ & $O((m+n+p)r^2)$ & $O((m+n+p)r^2)$ \\
    \quad Apply $\M{K}$ Completions  & $O((mn+mp+np)r)$ & -- \\
    \bottomrule
\end{tabular}
\end{center}
\caption{Comparison of per-iteration complexity of \cref{alg:newton_GLRT} with Gauss-Newton methods}
\label{tab:complexity}
\end{table}

\subsection{Generalization to d-way Case}
\label{sec:algorithm:dway}

While we detail Newton's algorithm for CP for the simpler 3-way case, we describe in this subsection how the derivations, computations, and conclusions generalize to the $d$-way case.
As usual, the notational complexity of the $d$-way case can obscure the key concepts.
We briefly introduce $d$-way notation and provide examples rather than comprehensive expressions to illustrate the ideas.
We consider $\Tn{X}$ to be a $d$-way data tensor with dimensions $n_1\times \cdots \times n_d$ and a rank-$r$ CP model with factor matrices $\left[\!\left[\M{U}{1},\dots,\M{U}{d}\right]\!\right]$.
We use $i_k$ to index dimension $n_k$ for $1\leq k\leq d$ and $\ell$ to index $r$ (as in the 3-way case), and we use MATLAB-style indexing with parentheses when subscript notation becomes overwhelming.
Note that the description of Gauss-Newton methods for the $d$-way case is given in \cite[Ch.~13]{TensorTextbook}.

The structure of the $\M{K}$ term of the $d$-way Hessian matches the one derived in \cref{sec:algorithm:der_hess}: we consider it as a symmetric block matrix with each block corresponding to the mixed partials of a pair of factor matrices.
This implies that the diagonal blocks are zeros and the $\binom{d}{2}=d(d-1)/2$ off-diagonal blocks are given by expressions that generalize \cref{lemma:newton3dmixedpartial}.
In the case of factor matrices $j$ and $k$, the corresponding off-diagonal block is given by $\M{K}{jk} = \normalfont{\texttt{blkdiag}}_{3} \left(\Tn{Z}{jk}\right)$ where $\Tn{Z}{jk}$ is the $n_j \times n_k \times r$ tensor computed from a partial MTTKRP involving the residual tensor and all factor matrices except $\M{U}{j}$ and $\M{U}{k}$.
Expressed elementwise for the case $j=1$ and $k=2$, for example, we have
$$\Tn{Z}{12}(i_1,i_2,\ell) = \sum_{i_3,\dots,i_d} \Tn{R}(i_1,\dots,i_d) \M{U}{3}(i_3,\ell) \cdots \M{U}{d}(i_d,\ell).$$

Applying the $\M{K}$ term of the $d$-way Hessian to a vector follows the pattern described in \cref{sec:algorithm:apply} and \cref{lem:Kvbar}: there are $d^2-d$ matrix-vector product terms to be computed involving a block of $\M{K}$ and a vectorization of a matrix with the dimensions of one of the two corresponding factor matrices.
This simplifies to an MTTKRP involving the residual tensor, all but two factor matrices from the Newton iterate, and one factor matrix from the arbitrary vector.
For a notationally convenient example, we have
\begin{align*}
\M{K}{12}\vc{\M[\bar]{U}{2}} &= \vc{\Tm{R}{1}\left(\M{U}{d}\krp \cdots \krp \M{U}{3} \krp \M[\bar]{U}{2}\right)}  \\
&= \vc{\M{U}{1}\left(\M{U}{d}'\M{U}{d}" \had \cdots \had \M{U}{3}'\M{U}{3}" \had \M{U}{2}'\M[\bar]{U}{2}"\right) - \Tm{X}{1}\left(\M{U}{d}\krp \cdots \krp \M{U}{3} \krp \M[\bar]{U}{2}\right)}. 
\end{align*}

As in the 3-way case, we must avoid explicitly computing the $d^2-d$ MTTKRPs involving either the residual tensor $\Tn{R}$ or the data tensor $\Tn{X}$ at every iteration of GLRT.
Instead, we precompute the $d(d-1)/2$ partial MTTKRPs involving the data tensor and all but two of the Newton iterate factor matrices.
Each of these intermediate tensors $\{\Tn{Y}\}$ are 3-way, with two dimensions corresponding to original tensor dimensions (that have not been contracted) and one dimension corresponding to the rank of the CP model.
Similar to the definition of $\Tn{Z}{12}$ above, we have
$$\Tn{Y}{12}(i_1,i_2,\ell) = \sum_{i_3,\dots,i_d} \Tn{X}(i_1,\dots,i_d) \M{U}{3}(i_3,\ell) \cdots \M{U}{d}(i_d,\ell).$$
Note that in the 3-way case, the partial MTTKRP involving all but two factor matrices simplifies to a TTM with one factor matrix.
We can also precompute the Gram matrices of the factor matrices of the Newton iterate (e.g., $\M{S}{k}=\M{U}{k}'\M{U}{k}"$).
Both sets of intermediates are re-used multiple times in evaluating the gradient and applying $\M{K}$.

In the 3-way case, the cost of the precomputations is dominated by the three TTMs, which is 50\% more than the cost of the two partial MTTKRPs required of ALS or Gauss-Newton methods (when memoization is used).
The key to the efficiency in the $d$-way case is to take advantage of memoization in the computation of the $\{\Tn{Y}\}$ tensors.
Despite the fact that the number of intermediates to be computed is $d(d-1)/2$, we can compute only 3 partial MTTKRPs involving the original data tensor using dimension tree memoization, and the rest of the computations are considerably cheaper \cite[Sec.~3.2.3]{MS22}.
The central idea is to choose 3 minimal overlapping subsets of modes such that every pair of modes appears in at least one subset.
This implies that the leading order cost of precomputations is still only 50\% more than that of ALS or Gauss-Newton methods in the $d$-way case.

Assuming a cubical $d$-way tensor $n_1 =\cdots =n_d$ and a rank-$r$ CP model, the complexity comparison given in \cref{tab:complexity} generalizes as follows.
In each outer iteration, the precomputation for Newton's method is dominated by $6n^dr$ compared to $4n^dr$ for Gauss-Newton.
The costs of gradient completions ($O(dn^2r)$), Gram matrices ($O(dnr^2)$), and rejected steps ($2n^dr$) are the same between methods. 
In each inner iteration, the cost of applying $\M{J}'\M{J}"$ shared by both methods is $O(d^2nr^2)$, and the added cost of applying $\M{K}$ is $O(d^2n^2r)$.
Because the cost of the precomputations in the outer iteration is the only cost exponential in $d$, we expect to be even more dominated by that step for $d>3$.
Thus, the estimate of Newton iterations costing 50\% more than Gauss-Newton iterations should be more accurate for larger $d$.

\section{Numerical Experiments}
\label{sec:experiments}
\newcommand{\drawmethod}[4]{%
  \foreach \t in {1,...,4}{%
    \addplot [#1, forget plot] table [x=iter, y=#2_\t_#3, col sep=space] {#4};%
  }%
  \addplot [#1] table [x=iter, y=#2_5_#3, col sep=space] {#4};%
}

\def\nanstr{NaN}
\newcommand{\drawfinalscore}[1]{%
  \pgfplotstableread{#1}\tltable
  \foreach \t in {1,...,5}{%
    \pgfplotstablegetelem{0}{TL_GN_\t_score}\of\tltable
    \let\tlscore\pgfplotsretval
    \pgfplotstableforeachcolumnelement{TL_GN_\t_err}\of\tltable\as\cell{%
      \ifx\cell\nanstr\else\xdef\tliter{\pgfplotstablerow}\fi}%
    \edef\temp{\noexpand\addplot [tlgncolor, only marks, mark=*, forget plot]
      coordinates {(\tliter, \tlscore)};}\temp
  }%
}

\newcommand{\drawswitch}[2]{%
  \pgfplotstableread[col sep=space]{#2}\swtable
  \foreach \t in {1,...,5}{%
    \pgfplotstablegetelem{0}{HYBRID_\t_gradnorm}\of\swtable
    \let\gzero\pgfplotsretval
    \addplot [only marks, mark=star, mark size=1.5pt, black, forget plot] table [x=iter,
      y expr={(\thisrow{HYBRID_\t_gradnorm} < 1e-4*\gzero && \prevrow{HYBRID_\t_gradnorm} >= 1e-4*\gzero && \thisrow{HYBRID_\t_gradnorm} == \thisrow{GN_\t_gradnorm}) ? \thisrow{HYBRID_\t_#1} : nan}] {\swtable};%
  }%
}

We implement \cref{alg:newton_GLRT} in MATLAB using Tensor Toolbox \cite{TensorToolbox} functions for many of the subroutines.
We compare against our implementation of Gauss-Newton, which is a modification of our Newton's method implementation, differing only in that it does not apply the $\M{K}(\vv)$ matrix within GLRT.
In addition, we consider a hybrid in which Gauss-Newton steps are taken until the norm of the gradient drops below a pre-specified limit, and thereafter Newton steps are taken. 
We also compare our Gauss-Newton implementation to that in Tensorlab~\cite{vervlietTensorlab30} to 
serve as a baseline, and also to ALS as implemented in Tensor Toolbox \cite{TensorToolbox}.
We measure progress in iterations rather than wall-clock time, as our MATLAB implementation has not been performance optimized.
We do not visualize rejected steps in the plots; as discussed in \cref{sec:algorithm:outline}, a rejected step has about 1/3rd the cost of an accepted step.
\Cref{tab:rejected} illustrates the number of rejected steps in one of the experiments as a representative example.

\subsection{Solvers}.
Each step of our Gauss-Newton (GN) and Newton solvers approximately minimizes a cubic-regularized model shown in \cref{eq:newton-model} using GLRT \cite{CartGoulToin22} with a block-diagonal preconditioner \cite{SVL13,VL19,TensorTextbook} and at most 100 inner iterations. 
The regularization weight starts at $\lambda=0.1$.
A step is rejected based on the ratio $\rho$ between the true and model decrease: a step is rejected if $\rho<10^{-8}$.
Accepted steps with $\rho<0.25$ multiply $\lambda$ by 3, and those with $\rho\ge0.75$ divide it by 2.
The inner tolerance follows the Eisenstat--Walker forcing terms~\cite{EW96}, with $\gamma=0.1$, $\eta_{\max}=10^{-2}$, and exponent $\alpha=1$ for GN and $\alpha=2$ for Newton.
The hybrid method runs Gauss-Newton until $\|\vg(\vv_k)\| < 10^{-4}\|\vg(\vv_0)\|$ and then switches permanently to Newton. These three methods stop when $\|\vg(\vv_k)\| < 3\times10^{-8}$ or when they reach the iteration limit.

`Tensorlab GN' is \texttt{cpd\_nls}, modified to stop on the gradient norm so that it matches our methods.\footnote{We copy Tensorlab's Gauss-Newton dogleg solver, \texttt{nls\_gndl}, and add a test at the start of each iteration that stops when the gradient norm at the current iterate falls below the tolerance; the solver is otherwise unchanged. We set the objective and step tolerances, \texttt{TolFun} and \texttt{TolX}, to zero, so those criteria stop the solver only if the change in the objective or the step is exactly zero.}
It uses the same gradient tolerance, iteration limit, and maximum number of CG iterations as our methods; all other options are left at their defaults.

The data tensor is scaled to unit Frobenius norm. In each trial we draw Gaussian random initial factors and scale them so that the initial model has the same norm as the data; this keeps the iterate away from the critical point at zero, where the gradient vanishes and the relative error is one. All methods in a trial start from the same initial guess. Trial $t$ uses random seed $t$, so different values of a swept parameter share the same initial tensor.

\subsection{Metrics}
We report the relative error $\|\Tn{X}-\Tn{M}\|/\|\Tn{X}\|$ and the gradient norm at each iteration, where $\Tn{X}$ is the original data, and $\Tn{M}$ is the CP model.
For synthetic tensors, the true factors are known, so we also measure a proxy for the forward error using the Tensor Toolbox \texttt{score} function, which matches computed components to true components using the cosine similarities between corresponding rank-one components \cite[Sec.~10.8]{TensorTextbook}.
A score of 1 indicates exact recovery.
When the model rank differs from the true rank, the extra components of the larger model are left unmatched. Because Tensorlab returns only the final factors, its score is shown only at the final iterate.
We stop on the gradient norm rather than the objective because the score can keep improving well after the relative error has leveled off (see \cref{fig:compare_noise_level}, for example).

Synthetic tensors are $200\times200\times200$, generated with Tensor Toolbox's \texttt{create\_problem}: the factor matrices have unit weights and columns with collinearity $c$, and noise is added as 
\begin{equation}
\label{eq:noise}
\Tn{X}=\Tn{M}+\frac{\eta}{\sqrt{1-\eta^2}}\cdot \|\Tn{M}\|\cdot \frac{\Tn{N}}{\|\Tn{N}\|}
\end{equation}
with $\Tn{N}$ Gaussian and $0\le\eta<1$. 
\footnote{
    We have modified \texttt{create\_problem} slightly.  Since $\Tn{N}$ is nearly orthogonal to $\Tn{M}$, $\|\Tn{X}\|^2\approx\|\Tn{M}\|^2/(1-\eta^2)$, so $\|\Tn{X}-\Tn{M}\|/\|\Tn{X}\|\approx\eta$: the true model has relative error close to $\eta$. 
    The standard \texttt{create\_problem} scaling, without the $\sqrt{1-\eta^2}$, yields a relative error of $\eta/\sqrt{1+\eta^2}$ for the true model, which differs noticeably for large $\eta$.
}

We demonstrate the performance of Newton's method on synthetic and real-world datasets.
As discussed in \cref{sec:globalization},
Gauss-Newton converges quickly when the residual at the solution is small, but its convergence slows as the residual grows.
Near a local minimizer, Newton's method retains fast convergence regardless of the residual.
Far from a minimizer, however, we observe that Newton's method can take more iterations than Gauss-Newton to find a neighborhood of a local minimum.
These findings motivate the hybrid method described at the start of Section~\ref{sec:experiments},
which starts with Gauss-Newton and switches to Newton's method when the gradient norm is sufficiently small.

Without noise, our synthetic tensors are exactly low rank, so a model of the true rank has zero residual at the solution.
We therefore introduce residual either by adding noise or by underestimating the rank, and we use high collinearity of the true factors to make the problem harder without introducing residual.
For visual clarity, not every method appears in every figure. In each figure, we show the trajectories for relative error, gradient norm, and score (if applicable) for each method and for five random trials.

\subsection{Varying Collinearity}
\begin{figure}[H]
\centering
\includegraphics{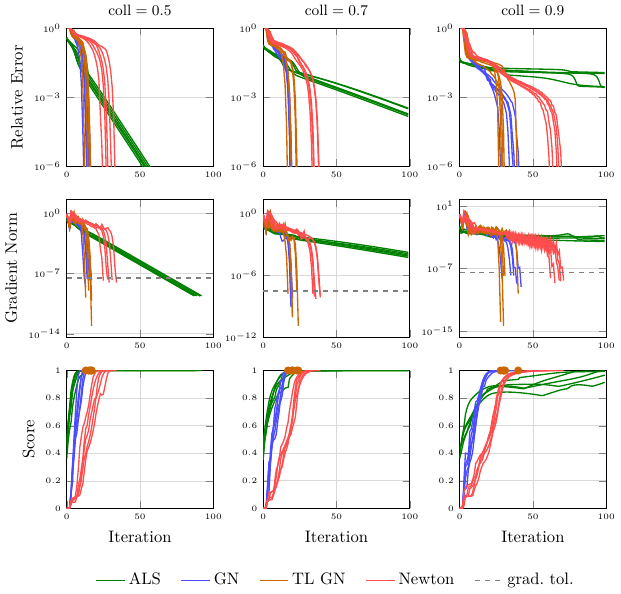}

\caption{Synthetic $200\times200\times200$ tensors with $r=r_{\text{true}}=10$ and $\eta=0$, for collinearity $c\in\{0.5,0.7,0.9\}$.
  Rows show relative error, gradient norm, and score; the dashed line marks the gradient tolerance.}
\label{fig:compare_collinearity}
\end{figure}

We first vary the collinearity $c$ of the true factors while keeping the residual at zero (\cref{fig:compare_collinearity}).
The model rank matches the true rank, $r=r_{\text{true}}=10$, and no noise is added, so a fitted model can attain zero relative error.
Each method runs for at most 100 iterations.

High collinearity is known to slow the convergence of ALS~\cite{mitchellSlowlyConvergingParafac1994,tomasiComparisonAlgorithmsFitting2006}.
As $c$ increases, ALS slows, and at $c=0.9$ it stalls: its relative error is high, its gradient norm barely decreases, and the score is sub-optimal.
In contrast, Gauss-Newton and Newton reach the gradient tolerance at every collinearity level and recover the true factors, with a score of 1.
Our Gauss-Newton implementation converges in a similar number of iterations as the Tensorlab implementation, so we will use it as a baseline to compare against Newton.

Newton takes approximately twice as many iterations as Gauss-Newton at every collinearity level, although both achieve superlinear convergence eventually.
Since the residual vanishes at the solution, so does $\M{K}$, and the two methods take nearly the same steps once they are close to a minimizer.
The difference therefore comes from the early iterations: far from a minimizer, $\M{K}$ is not negligible, and Newton wanders before it reaches the region where it converges quickly.

\subsection{Varying Noise}
\begin{figure}[H]
\centering
\includegraphics{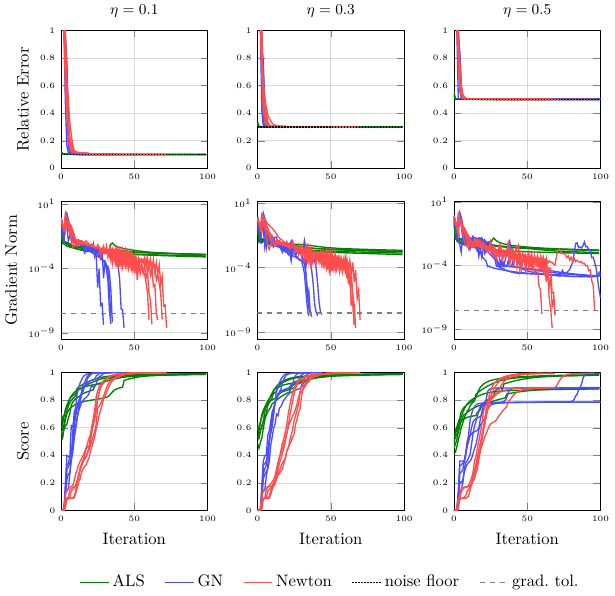}

\caption{Synthetic $200\times200\times200$ tensors with $r=r_{\text{true}}=10$ and $c=0.9$, for noise parameter $\eta\in\{0.1,0.3,0.5\}$.
  Rows show relative error, gradient norm, and score; the dotted line marks the relative error of the true model, approximately $\eta$, and the dashed line marks the gradient tolerance.}
\label{fig:compare_noise_level}
\end{figure}

Next, we fix collinearity at $c=0.9$ and rank $r=r_{\text{true}}=10$ and vary the noise parameter $\eta\in\{0.1,0.3,0.5\}$ from \eqref{eq:noise}, so that the true model has relative error approximately $\eta$ (\cref{fig:compare_noise_level}).
Because the data is otherwise exactly low rank, a large amount of noise is needed before the residual noticeably affects Gauss-Newton.
Each method runs for at most 100 iterations.

For $\eta=0.1$ and $\eta=0.3$, both Gauss-Newton and Newton converge in every trial, with Newton again taking roughly twice as many iterations.
At $\eta=0.5$, however, Gauss-Newton does not reach the tolerance in any trial.
Newton reaches the tolerance in every trial in a similar number of iterations as for the smaller noise levels, converging superlinearly once it is close to a minimizer.

At $\eta=0.5$, Newton attains a score of about 0.99 in four of the five trials, while Gauss-Newton reaches the iteration limit with a score below 0.9 in four of the five trials.
This high level of noise demonstrates Newton's robustness, and these results show that it is still able to recover the true signal.

\subsection{Varying Rank}
\begin{figure}[H]
\centering
\includegraphics{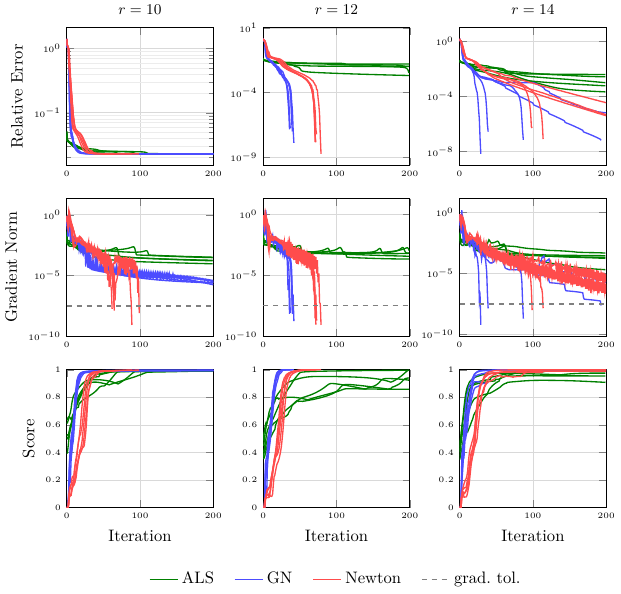}

\caption{Synthetic $200\times200\times200$ tensors with $r_{\text{true}}=12$, $c=0.9$, and $\eta=0$, for model rank $r\in\{10,12,14\}$ (underestimated, correct, overestimated).
  Rows show relative error, gradient norm, and score; the dashed line marks the gradient tolerance.}
\label{fig:compare_ranks}
\end{figure}

In practice, the true CP rank is not known in advance, so a model may underestimate or overestimate it.
We fix $r_{\text{true}}=12$, $c=0.9$, and $\eta=0$, and fit models of rank $r\in\{10,12,14\}$ (\cref{fig:compare_ranks}).
Each method runs for at most 200 iterations in this experiment.

Underestimating the rank leaves the two unmodeled components in the residual.
Although the resulting relative error is only about 0.02, Gauss-Newton does not reach the tolerance in any trial, whereas Newton converges in every trial.

When the rank is correct, the residual is zero, and the results resemble those of the collinearity experiment: Newton takes roughly twice as many iterations as Gauss-Newton.

Overestimating the rank also allows zero residual, but the solution is no longer unique, and convergence is less consistent for every method.
Within 200 iterations, Gauss-Newton converges in four trials, and Newton converges in two.
Even when a method converges, its score can be less than 1: each true component is matched to a single computed component and any extra components are ignored, so a true component that is split between two computed components is only partially recovered.

\subsection{Motivation for the Hybrid Method}
\begin{figure}[H]
\centering
\includegraphics{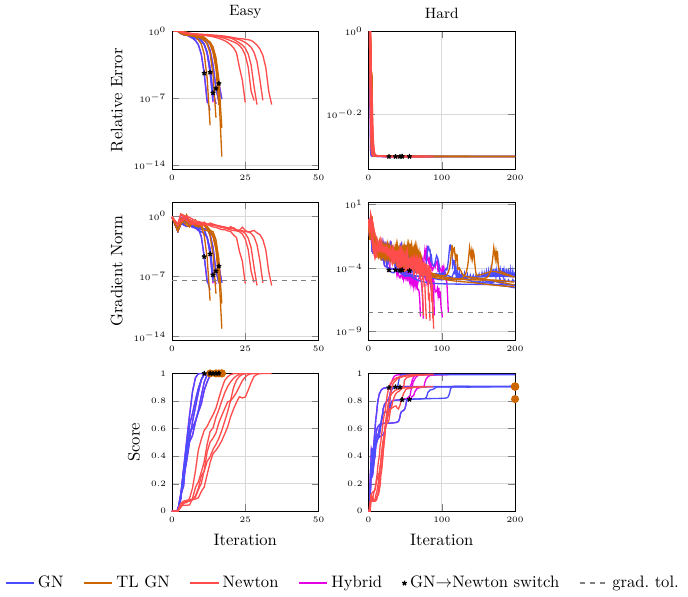}

\caption{Synthetic $200\times200\times200$ tensors.
  Left (Easy): $r=r_{\text{true}}=10$, $c=0.5$, and $\eta=0$.
  Right (Hard):$r=14$, $r_{\text{true}}=12$, $c=0.9$, and $\eta=0.5$.
  Rows show relative error, gradient norm, and score; the dashed line marks the gradient tolerance.}
\label{fig:compare_easy_vs_hard}
\end{figure}

We conclude the synthetic experiments by comparing the gradient-based methods on two illustrative problems (\cref{fig:compare_easy_vs_hard}) that motivate the use of a hybrid method. We contrast two cases: `easy' and `hard'.
The easy problem has low collinearity, the correct rank, and no noise. The hard problem combines the difficulties of the previous experiments: high collinearity, an overestimated rank, and high noise.

On the easy problem, Gauss-Newton and Tensorlab GN converge quickly. Note that the x-axis is truncated to 50 iterations for the easy problem to better highlight the convergence pattern. Newton takes about twice as many iterations as Gauss-Newton similarly to the collinearity experiment.
The hybrid method reaches the switching threshold only one iteration before it converges, so it follows Gauss-Newton almost exactly. All methods achieve a perfect score of 1, exactly recovering the true factors.

On the hard problem, neither Gauss-Newton implementation reaches the tolerance in any trial, so this failure is not specific to our implementation.
Newton and the hybrid method converge in every trial and also reach a high score in all trials, while Gauss-Newton and Tensorlab GN mostly stop near 0.9.
For this experiment, we also report the number of accepted and rejected steps taken by each method in \cref{tab:rejected}. 
On the hard problem, the hybrid method rejects roughly 20 more steps per trial than Gauss-Newton but takes roughly 100 fewer accepted steps. Since a rejected step costs about a third of an accepted one (see \cref{sec:algorithm:outline}), counting rejected steps does not change the conclusion.

The hybrid method therefore costs almost nothing when Gauss-Newton converges quickly, and converges like Newton when Gauss-Newton fails, without knowing in advance which kind of problem it faces.

\begin{table}[ht]
    \centering
    \begin{tabular}{lcccc}
        \toprule
        & \multicolumn{2}{c}{Easy} & \multicolumn{2}{c}{Hard} \\
        \cmidrule(lr){2-3} \cmidrule(lr){4-5}
        Method & Accepted & Rejected & Accepted & Rejected \\
        \midrule
        Gauss-Newton  & 13--17 & 4--7  & 200     & 7--13  \\
        Tensorlab GN  & 13--17 & 2--4  & 200     & 20--26 \\
        Newton   & 25--34 & 9--15 & 76--89  & 30--35 \\
        Hybrid  & 12--17 & 4--7  & 71--109 & 20--32 \\
        \bottomrule
    \end{tabular}
    \caption{Range over 5 trials of the number of accepted and rejected steps for each method.
    The iteration limit counts accepted steps only.}
    \label{tab:rejected}
\end{table}

\subsection{Gas Sensor Array}
\begin{figure}[H]
\centering
\includegraphics{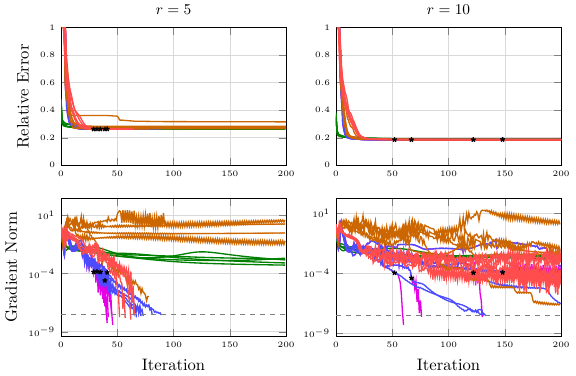}

\includegraphics{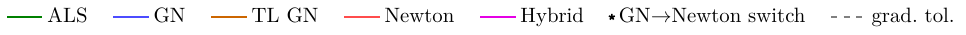}

\caption{Gas sensor array tensor ($1036\times72\times900$), for $r\in\{5,10\}$.
  Rows show relative error and gradient norm; the dashed line marks the gradient tolerance.
  The true factors are unknown, so no score is shown.}
\label{fig:gas_results}
\end{figure}

We next consider the gas sensor array dataset~\cite{vergaraGasSensorArrays2013}, which has been used to benchmark CP methods in~\cite{vervlietRandomizedBlockSampling2016,battaglinoPracticalRandomizedCP2018}.
We follow the preprocessing of~\cite{vervlietRandomizedBlockSampling2016}, except that we keep every 25th time sample to reduce running time, giving a $1036\times72\times900$ tensor (time $\times$ sensor $\times$ experiment).\footnote{The preprocessing script of~\cite{vervlietRandomizedBlockSampling2016} discards one file that fails to parse because of a stray non-ASCII character, leaving 899 experiments. We remove the character and keep the file.}
We fit $r=10$, as in~\cite{vervlietRandomizedBlockSampling2016,battaglinoPracticalRandomizedCP2018}, and also $r=5$, for which convergence is more regular (\cref{fig:gas_results}).
Each method runs for at most 200 iterations.
Since the true factors are unknown, we report only relative error and gradient norm.

Unlike our synthetic tensors, this data is not close to low rank: the computed minimizers have relative errors of about 0.26 for $r=5$ and 0.19 for $r=10$.
For $r=5$, Gauss-Newton, Newton, and the hybrid method converge in every trial, but in contrast to the zero-residual synthetic problems, Newton converges in fewer iterations than Gauss-Newton.
The hybrid method converges faster still, while ALS and Tensorlab GN do not reach the gradient tolerance in any trial.
For $r=10$, no method converges consistently: the hybrid method converges in three trials, Gauss-Newton in two, and Newton in none.
In the trials where both converge, the hybrid method takes about half as many iterations as Gauss-Newton.

\subsection{Monkey BMI}
\begin{figure}[H]
\centering
\includegraphics{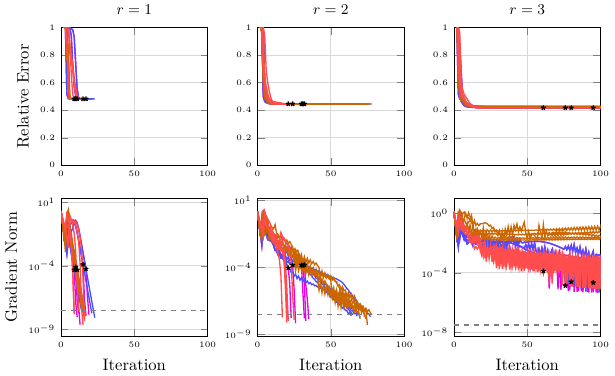}

\includegraphics{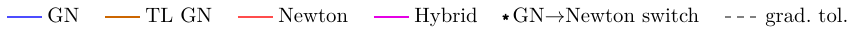}

\caption{Monkey BMI tensor ($43\times200\times88$), for $r\in\{1,2,3\}$.
  Rows show relative error and gradient norm; the dashed line marks the gradient tolerance.
  The true factors are unknown, so no score is shown, and only the first 100 of 200 iterations are shown.}
\label{fig:monkey_results}
\end{figure}

Finally, we consider the Monkey BMI dataset from~\cite{TensorTextbook}, originally collected in~\cite{vyasNeuralPopulationDynamics2018}, a $43\times200\times88$ tensor (neuron $\times$ time $\times$ trial).
In~\cite{TensorTextbook}, the rank is chosen where the relative error stops significantly improving, and a nonnegative model of rank 10 is fit.
Since our goal is instead to study convergence in the gradient norm, we fit unconstrained models of rank $r\in\{1,2,3\}$ (\cref{fig:monkey_results}). The convergence behavior for $r=4$ to 10 is similar to that for $r=3$ and is not shown.
The relative errors of the computed minimizers are about 0.48, 0.45, and 0.42, respectively, roughly in line with our $\eta = 0.5$ synthetic experiment.

For $r=1$, every method converges within 25 iterations.
For $r=2$, Newton converges in about a third as many iterations as Gauss-Newton, and the hybrid method is only slightly slower than Newton.
For $r=3$, no method reaches the tolerance within the iteration limit, although Gauss-Newton, Newton, and the hybrid method all reach the same relative error.

\section{Conclusion}

We have presented an efficient way to apply the true Hessian of the CP model's objective function to a vector. This allows us to employ a safeguarded Newton method to solve the CP optimization problem, as this can converge much faster than currently used CP algorithms such as ALS and Gauss-Newton (when the residual is large). We have shown that the per-outer-iteration cost of a applying the Newton Hessian is roughly $50\%$ higher than that of Gauss-Newton for large tensors.

Gauss-Newton and Newton models for CP have intrinsic issues arising from the ambiguity of ordering and scaling the columns in the factor matrices, resulting in rank-deficient Hessians.
An adaptive cubic-regularization technique allowed both methods to overcome the rank deficiency issue and to globalize them, that is, to ensure that the generated iterates converge from any starting point.

Numerical experiments on a variety of synthetic problems and real datasets demonstrated that the asymptotic convergence of Gauss-Newton depends on the difficulty of the problem at hand.
The convergence rate of Gauss-Newton ranged from superlinear on easy problems to linear on mild to hard problems (even stagnation in some cases), while Newton-based methods exhibit asymptotic superlinear convergence. 
In this case, ``hard'' problems refer to inconsistent problems, large nonzero residual problems and/or highly ill-conditioned factor matrices, which frequently occur in practice.
Since Newton convergence was often slow at the start of the minimization, and noticeably slower than that of Gauss-Newton, a hybrid method that starts with a Gauss-Newton model and switches to a Newton model exhibits the best features of both methods, and this is what we recommend practitioners use. The main caveat of the hybrid method is that the threshold for switching from Gauss-Newton to Newton may need tuning.

With savings beyond a factor of two iterations and success where Gauss-Newton struggles to converge, at the cost of $\sim$50\% extra in the last few iterations, we recommend the hybrid method as the default for Hessian-based CP fitting.
While our complexity analysis supports using the number of iterations as a proxy for time, the next step is to demonstrate that the hybrid method also reduces time compared to the alternatives.
Future work will replace our MATLAB implementation with optimized computational kernels, including techniques for memoization in the $d$-way case discussed in \cref{sec:algorithm:dway}.

While the equilibration technique given by \cref{eq:equilibrate} has
proved effective in practice, it might be better to try to enforce 
normalization directly as part of the minimization problem. For example,
in the 3-way case, we currently employ the unweighted rank-$r$ Kruskal tensor
with all the ambiguity that it implies. Another possibility is to use
the weighted variant
$\sum_{i=1}^r \lambda_i \, \va_i \top \vb_i \top \vcc_i,$
where now we require that 
$\|\va_i\| = \|\vb_i\| = \|\vcc_i\| = 1$
and $\lambda_i \geq 0$.
These norm restrictions say that the unknowns in 
$(\va_i,\vb_i,\vcc_i)$ lie on a Riemannian manifold, and there are
manifold optimization methods \cite{Boum23} that impose such constraints 
explicitly, thereby removing local ambiguities. We are currently 
considering the application of such methods to this alternative formulation,
and note that the required Jacobian and Hessian products we have developed
are easily modified for such a Riemannian derivative case.

\section*{Acknowledgements}
This material is based upon work supported by the US Department of Energy, Office of Science, Advanced Scientific Computing Research program under awards DE-SC0023296 and DE-SC0025394.
Al Daas and Gould are supported by the Ada Lovelace Centre Programme at the Scientific Computing Department, STFC.

During the preparation of this work, the authors used Claude to assist with the following tasks: writing the MATLAB code for our experiments, drafting both the Hessian derivations and the discussion of each experiment, creating the TikZ code for each figure in the manuscript, and checking for mathematical and notational consistency. Afterwards, the authors reviewed and edited the content as needed and take full responsibility for the content of the published article.

\bibliographystyle{plainurl}
\bibliography{refs}

\end{document}